\documentclass[superscriptaddress,amssymb,twocolumn,aps]{revtex4-1}

\usepackage{natbib,graphicx,dcolumn,bm,amsmath,color,bm,subfigure}

\usepackage{soul,xcolor}
\setstcolor{red}

\newcommand{\eq}[1]{Eq.~(\ref{#1})}

\newcommand{\fig}[1]{Fig.~\ref{#1}}

\definecolor{amber}{rgb}{1.0, 0.75, 0.0}

\newcommand{\eeq}{ \end{equation} }
\newcommand{\beq}{ \begin{equation} }

\newcommand{\eea}{ \end{align} }
\newcommand{\bea}{ \begin{align} }

\newcommand{\oma}{{\bf \hat{u}}}

\newcommand{\bhu}{ {\bf \hat{u}} }

\newcommand{\pp}{{\mathcal P}_{2}}

\newcommand{\bn}{ {\bf \hat{n}} }

\newcommand{\ellp}{\ell^{\prime}}

\newcommand{\red}[1]{ { \color{red} #1 } }

\begin{document}

\title{Second-virial theory for shape-persistent living polymers templated by discs}

\author{M. Torres L\'{a}zaro}
\affiliation{Laboratoire de Physique des Solides - UMR 8502, CNRS, Universit\'{e} Paris-Saclay, 91405 Orsay, France}
\author{R. Aliabadi}
\affiliation{Department of Physics, Faculty of Science,  Fasa University, 74617-81189 Fasa, Iran}
\author{H. H. Wensink}
\email{Corresponding author \\ E-mail: rik.wensink@universite-paris-saclay.fr}
\affiliation{Laboratoire de Physique des Solides - UMR 8502, CNRS,  Universit\'{e} Paris-Saclay, 91405 Orsay, France}

\date{\today}

\begin{abstract}

Living polymers composed of non-covalently bonded  building blocks with weak backbone flexibility may self-assemble into thermoresponsive lyotropic liquid crystals. We demonstrate that the reversible polymer assembly and phase behavior  can be  controlled by the addition of (non-adsorbing) rigid colloidal discs which act as an entropic reorienting ``template"  onto the supramolecular polymers. Using a particle-based second-virial theory that correlates the various entropies associated with the polymers and discs, we demonstrate that small fractions of discotic additives promote the formation of a polymer nematic phase. At larger disc concentrations, however, the phase is disrupted by collective disc alignment in favor of a discotic nematic fluid in which the polymers are dispersed anti-nematically. We show that the anti-nematic arrangement of the polymers generates a non-exponential molecular-weight distribution and stimulates the formation of oligomeric species. At sufficient concentrations the discs  facilitate a liquid-liquid phase separation which can be brought into simultaneously coexistence with the two fractionated nematic phases, providing evidence for a four-fluid coexistence in reversible shape-dissimilar hard-core mixtures without cohesive interparticle forces.  We  stipulate the conditions under which such a phenomenon could be found in experiment.
\end{abstract}

\maketitle

\section{Introduction}

 Supramolecular "living" polymers are composed of  aggregating building blocks that are joined together via non-covalent bonds.  The polymers can break and recombine reversibly as the typical attraction energy between monomers is comparable to the thermal energy \cite{cates87,cates88}.  Elementary (Boltzmann) statistical mechanics then tells us that the polymers must be in equilibrium with their molecular weight distribution which emerges from a balance between the association energy and mixing entropy of the polymers. This results in a wide range of different polymeric species with an exponential size distribution whose shape is governed primarily by temperature and  monomer concentration. Reversible polymers are thus distinctly different from usual ``quenched" polymers whose molecular weight distribution is fixed primarily by the conditions present during the synthesis process.
 
  Reversible association is ubiquitous in soft matter. Examples include the formation of various types of micellar structures from block-copolymers \cite{riess2003,blanazs2009}, hierarchical self-assembly of  short-fragment DNA  \cite{demichele2012,demichele2016}, chromonic mesophases  \cite{lydon2010,tamchang2008}  composed of non-covalently stacked sheetlike macromolecules, and the assembly of amyloid fibrils from individual proteins \cite{knowles2011}. Microtubules, actin and other biofilaments provide essential mechanical functions in the cell and consist of dynamically organizing molecular units that self-organize into highly interconnected structures \cite{fuchs1998}.

A particularly interesting case arises when the monomers associate into shape-persistent, directed polymers \cite{gittes1993}. Interpolymer correlations then become strongly orientation-dependent and may drive the formation of liquid crystals.  Spontaneous  formation of lyotropic liquid crystals has been observed, for example, in long worm-like micelles under shear  \cite{berret1994}, oligomeric DNA \cite{nakata2007} and chromonics \cite{lydon2010}. When the monomer concentration exceeds a critical value, the polymers grow into strongly elongated aggregates  and an (isotropic) fluid of randomly oriented polymers may spontaneously align into, for instance, a nematic liquid crystal characterized by long-range orientational correlations without structural periodicity \cite{gennes-prost}.  While aggregation-driven nematization has been contemplated also for thermotropic systems  \cite{matsuyama1998}, our current focus  is  on lyotropic systems composed of rigid polymers suspended in a fluid host medium, where the isotropic-nematic phase transition can be rationalized on purely entropic grounds in terms of a gain of volume-exclusion entropy upon alignment at the expense of orientational entropy \cite{Onsager, odijkoverview,Vroege92}. However, this argument becomes more convoluted in the case  of directed, reversible polymers where the trade-off between these two entropic contributions becomes connected with a simultaneous maximisation of the mixing entropy and the number of monomer-monomer linkages. In particular, the  coupling between orientational order and polymer growth turns out to be a very important one; collective alignment leads to longer polymers, which tend to align even more strongly thus stimulating even further growth \cite{vdschoot1994la}. More recent simulation studies have basically corroborated this scenario \cite{kindt2001,kuriabova2010,nguyen2014}.  

\begin{figure*}
  \includegraphics[width=\textwidth]{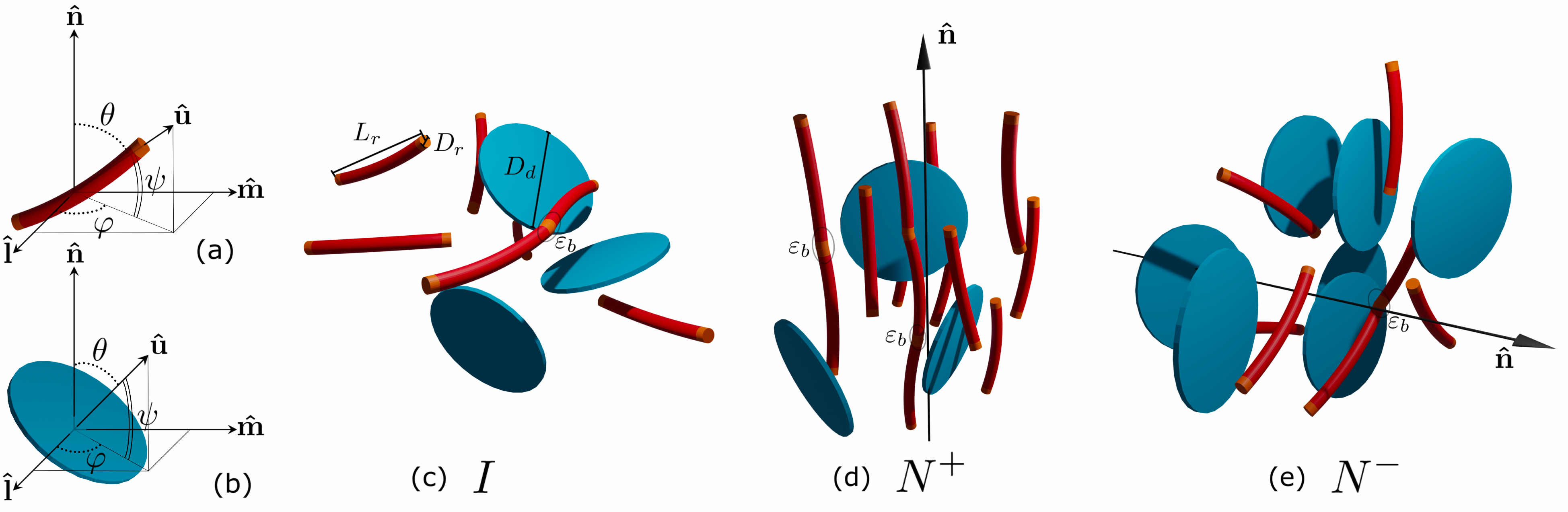}
  \caption{Schematic representation of the various liquid crystal phases emerging for discs mixed with polymerizing rods: (a) - and (b) - Principal angles describing the orientation $\bhu$ of a single rod monomer - and disc - with respect to the molecular director $\bn$  with $\theta$ denoting the polar angle, $\varphi$ the azimuthal angle and $\psi = \frac{\pi}{2} - \theta $ the meridional angle.  (c) Isotropic phase. (d) polymer uniaxial nematic phase $N^+$. (e) discotic uniaxial nematic phase $N^-$ in which the reversibly polymerizing rods are dispersed {\em anti-nematically}.}
  \label{fig:cartoon}
\end{figure*}

An intriguing question in relation to the above is the following:  Can the hierarchical organization of reversible polymers be controlled by the addition of non-adsorbing shape-dissimilar components that affect the way they align? Indeed, for chromonics it is known that the presence of additives can bring about condensation or reorientation of the reversible stacks, thereby changing their phase behavior through subtle modifications of the system entropy \cite{tortora2010}.  Recent experiments on  clay nanosheets mixed with reversibly polymerizing tubuline rods have demonstrated that these mixtures remain stable against flocculation and provides a testbed for exploring entropy-driven phase behavior of biopolymer-platelet mixtures \cite{kato2018}.
Furthermore, it is well established that mixing prolate (rod-shaped) colloids with their  oblate  counterparts  generates a strong coupling between the  orientations of both components leading to organizations with mixed nematic and anti-nematic symmetries. Numerous theoretical studies starting with the early work of Alben \cite{alben1973} have attempted to rationalize the intricate isotropic-nematic  phase behavior of these mixtures  placing particular emphasis on stabilizing the highly sought-after biaxial nematic phase in which both components are aligned along mutually perpendicular directions thus generating a fluid with an orthorhombic ($D_{2h}$) symmetry  \cite{stroobants1984,campallenbolhuisfrenkel,sokolova1997,vanakaras1998,vanakaras2001, matsuda2003,jacksonbiaxrev,varga2002,galindo2,wensinkrodplate,wensinkbiaxial}. Similar kinds of  anti-nematic or biaxial symmetries could arise when dispersing rod-shaped colloids in a thermotropic liquid crystal under appropriate anchoring conditions \cite{matsuyama2010,mundoor2018}. Anti-nematic order has been shown to naturally emerge in porous smectic structures of  shape-persistent nanorings \cite{avendano2016,wensinkavendano2016} or may be realized with the help of external electromagnetic fields as was demonstrated for clay nanosheets \cite{dozov2011} and for discs in the presence of associating magnetic beads \cite{perouklapp2020}.  In this study we wish to build upon the preceding concepts and explore hierarchical self-organization of reversible polymers in the presence of disc-shaped particles. An example of colloidal discs that could be envisaged are clay nanosheets that consist of nanometer-thick discotic particles with a very high diameter-to-thickness ratio. These particles find widespread use in industrial soft matter and are at the basis of many colloidal-polymer composite materials \cite{balazs1998,ginzburg2000}. The clay sheets on their own, provided they do no gelate in crowded conditions,  have a natural tendency to align and form various types of liquid crystals, including nematic phases \cite{kooij1998,gabriel2005,michot2006,paineau_jpcb2009}. When mixed with  reversibly polymerizing components in the absence of strong disc-polymer attractions,  the discs not only induce orientational "templating" of the supramolecular polymers \cite{asdonk2017}, they also influence the mixing entropy of the system which must have  consequences for polymer growth and  phase behavior \cite{taylor1993,vdschoot1994epl}. It is precisely these combined entropic effects that we wish to examine more closely in this work. To this end, we formulate a simple model (Section II) that we subsequently cast into  a  particle-based theory (Section III) that features reversible association and accounts for all relevant entropic contributions on the approximate second-virial level. The orientation degrees of freedom of the species are treated using a number of simplified variational approaches that render our theory algebraically manageable.    We stress that our primary attention in this work goes to mixed-shape nematic phases and we do not consider partially crystallized states that may become stable at elevated packing conditions where our theoretical approach is no longer applicable.

Our study broadly falls into two parts. In the first part (Section IV) we explore the molecular weight  distribution in mixtures in which the polymers are organized either nematically or {\em anti-nematically}. The latter state can be realized at elevated disc concentrations where  correlations between the discs are strong enough to generate nematic order of the discotic subsystem which in turn, enforces the supramolecular rods to align perpendicular to the discotic director in such a way that the overall system retains its  uniaxial $D_{\infty  h}$ point group symmetry (\fig{fig:cartoon}(e)). Whereas reversible polymers in a conventional nematic organization are distributed along a near-exponential form with minor non-exponential corrections at short lengths \cite{kuriabova2010},  we argue that {\em anti-nematic} living polymers may, under certain conditions, exhibit a strong non-exponential weight distribution with the most-probable polymer size being oligomeric rather than monomeric. 

In the second part of the manuscript (Section V and VI) we explore the isotropic-nematic phase behavior of the mixed systems by focusing on the uniaxial nematic phases, which seems to be the prevailing nematic symmetry for strongly shape-dissimilar mixtures \cite{wensinkrodplate,campallenbolhuisfrenkel,varga2002,matsuda2003,jacksonbiax}. Our theoretical model is generic and should be applicable to a wide range of different monomer-disc size ratios and temperatures. We discuss the key features for a few exemplary mixtures. One of them is a distinct azeotrope that develops for the isotropic-polymer nematic coexistence, suggesting a strong orientational templating effect imparted by volume-excluded interactions between the polymers and the discs.  Furthermore,  under certain disc-monomer size constraints,  a remarkable four-phase equilibria appears involving a simultaneous coexistence of isotropic gas and liquid phases along with two fractionated uniaxial nematic phases. In Section VII we discuss our findings in relation to recent colloid-polymer models where similar multiphase equilibria have been reported. We end this work with formulating the main conclusions along with some perspectives for further research in Section VIII.

\section{Model}

In this study, we focus on mixtures of tip-associating rod-shaped monomers with limited backbone flexibility  mixed with rigid discs. An overview of the basic particle shapes is given in \fig{fig:cartoon}. We assume that each rod monomer is equipped with identical attractive patches at either tip such that each rod end can only form a single bond  with an adjacent rod tip producing a linear polymer. The rods do not associate into multi-armed or ring-shaped polymers. We further assume that all species retain their basic fluid order such that the respective density distributions remain uniform in positional space (but not necessarily in orientational phase space). We do not account for the possibility of hexagonal columnar phases formed by (pure) polymers at high monomer concentration and low temperature combined with elevated polymer backbone flexibility  \cite{taylorherzfeld1991,schoot1996}. In fact, discs too may form columnar structures at packing fraction exceeding typically 40 \% \cite{frenkelcol1989,veerman1992,kooij_nature2000} which goes beyond the concentration range we consider relevant here.  Interactions between the polymer segments and the discs are assumed to be purely hard with the only energy scale featuring in the model being the non-covalent bond energy $\varepsilon_{b}$ between the monomers.

 Contrary to previous modelling studies of rod-discs mixture we focus here solely on uniaxial nematic phases and ignore the possibility of biaxial order in which both components align along mutually perpendicular directors. Our focus is motivated by  the fact that we expect excluded-volume interactions between the polymers and the discs, which are the principal entropic forces behind generating nematic order \cite{Onsager}, to be too disparate to guarantee such orthorhombic nematic symmetry to be stable. Previous theoretical studies \cite{jacksonbiaxrev,varga2002,jacksonbiax,wensinkrodplate,wensinkbiaxial} as well as experiments \cite{kooijlangmuir2000,kooijprl2000,woolston2015} and simulations \cite{campallenbolhuisfrenkel,galindo1,galindo2} on mixed-shape colloids  suggest that strongly unequal excluded volumes indeed favour demixing into strongly fractionated uniaxial nematic phases.  In view of the basic symmetry difference  between the linear polymer and disc, we  then anticipate a rod-based uniaxial phase (denoted $N^{+}$,  \fig{fig:cartoon}(d)) in which the discs are distributed  anti-nematically throughout the uniaxial matrix. Conversely, when the discs outnumber the polymers,  a disc-based uniaxial nematic ($N^{-}$,   \fig{fig:cartoon}(e)) is formed in which the aggregating rods adopt anti-nematic order. The onset of biaxial order emerging from these uniaxial reference phases can be estimated from a simple bifurcation analysis discussed in Appendix B.

\section{Second-virial Theory for Reversible Polymers mixed with rigid discs}

We start with formulating the free energy per unit volume $V$ of a mixture of discs with density $\rho_{d}(\oma) $ and reversibly polymerizing rods. We define $\rho_{r}(\ell, \oma)$ as the number density of monomer segments  aggregated into a polymeric rod with contour length $\ell L$  and orientation described by unit vector $\oma$.  The aggregation number or polymerization degree is specified by the index $\ell =1,2,3,...$.  Let us write the free energy per unit volume of the mixture as follows \cite{kuriabova2010,wensink_mm2019}:
 \begin{align}
& \frac{  F}{ V}  \sim  \sum_{\ell} \int  d \oma   \left [ \ln \left (4 \pi \Lambda_{r} \rho_{r} (\ell, \oma) \ell^{-1}  \right ) - 1 \right ] \ell^{-1} \rho_{r} (\ell, \oma) \nonumber \\ 
& +     \int  d \oma   \left [ \ln \left ( 4 \pi \Lambda_{d} \rho_{d} ( \oma)   \right ) - 1 \right ]  \rho_{d} (\oma)  + \frac{F_{as}}{V} +  \frac{F_{wlc}}{V} + \frac{F_{ex}}{V} \nonumber \\ 
 \label{free}
\end{align}
Without loss of generality,  all energies are implicitly expressed in units of thermal energy $k_{B}T$ (with $k_{B}$ Boltzmann's constant and $T$ temperature). Furthermore, $\Lambda_{r/d}$ are the thermal volumes of the species  which are immaterial for the thermodynamic properties we are about to explore. The factor $4 \pi$ is included for convenience and equals the unit sphere surface representing the orientational phase space. 
The total rod monomer concentration $\rho_{r0}$ is a conserved quantity so that $\rho_{r0} = \sum_{\ell} \int  d \oma \rho_{r}(\ell, \oma )$. Likewise, $\rho_{d0} = \int d \oma \rho_{d} (\oma)$ represents the number  density of discs.
The first two terms are related to the ideal gas or mixing entropy and describe the ideal translation and orientational entropy of each polymer  and disc, respectively.
The third contribution in \eq{free} represents an association energy that drives end-to-end aggregation of the  monomer segments. It reads: 
 \beq
\frac{F_{as}}{V}  =  \varepsilon_{b}  \sum_{\ell} \int d \oma \ell^{-1} \rho_{r}(\ell, \oma) (\ell -1 ) 
 \label{fas} 
 \eeq
 The free energy per unit volume arising from the polymerized rod segments follows  from the bond potential $\varepsilon_{b}$ between two adjacent rod segments and the number density  $\rho_{a}(\ell, \oma) = (1/\ell) \rho_{r}(\ell, \oma)$ of polymers with aggregation number $\ell  $  each containing  $\ell -1$ bonds. Being normalized to the thermal energy the potential $\varepsilon_{b}$ serves as an {\em effective} temperature scale. At strongly reduced temperature ($\varepsilon_{b} \ll 0$) the association energy is minimised when all monomers  join together into a single long  polymer, while at high temperature ($\varepsilon_{b} \gg 0$) polymerization is strongly suppressed. If $-\varepsilon_{b}$ is of the order of the thermal energy $k_{B}T$, the single chain configuration is highly unfavorable in view of the mixing entropy that favors a broad distribution of aggregates with strongly disperse contour lengths. This we will explore more systematically in Section IV.  

\subsection{Backbone flexibility}

The second last term in \eq{free} represents the effect of polymer flexibility through a correction to the  original orientational entropy (first term in \eq{free}) that accounts for the internal configurations of a so-called worm-like chain \cite{Vroege92}. This leads to a  strongly non-linear term with respect to the segment density \cite{khokhlov82,kuriabova2010}: 
\beq
  \frac{F_{wlc}}{V} = - \frac{2L_{r}}{3\ell_{p}} \sum_{\ell} \int d \oma [ \rho_{r}(\ell, \oma)]^{1/2} \nabla^{2}  [ \rho_{r}(\ell, \oma)]^{1/2}
  \label{wlc}
\eeq
where $\nabla^{2} $ denotes the Laplace operator on the unit sphere. 
The persistence length $\ell_{p}$ measures the  typical length scale over which local orientational fluctuations of the segments are correlated.  In our model we assume that the rod segments are only slightly flexible \cite{khokhlov82}  so that $\ell_{p} \gg \ell$ suggesting that the main orientational entropy stems from the rigid body contribution that is subsumed into the ideal gas term in \eq{free}. The worm-like chain correction   
vanishes in the somewhat unnatural situation where all polymers, irrespective of their contour length, are perfectly rigid and the persistence length tends to infinity ($\ell_p \rightarrow \infty$).

\subsection{Excluded-volume entropy}

The last contribution in \eq{free} is the excess free energy that incorporates all excluded-volume driven interactions between the stiff polymers and discs. Assuming all interactions to be strictly hard, we write following Ref. \cite{stroobants1984} :
\begin{align}
& \frac{F_{ex}}{ V}  =  \frac{1}{2}   \sum_{\ell, \ell^{\prime}}  \iint d  \oma  d \oma^{\prime} \rho_{r}(\ell, \oma)  \rho_{r}(\ell^{\prime}, \oma^{\prime})  2  L_{r}^{2} D_{r}  | \sin \gamma |  \nonumber \\
& +    \sum_{\ell}  \iint d  \oma  d \oma^{\prime} \rho_{r}(\ell, \oma) \rho_{d}( \oma^{\prime})  \frac{\pi}{4}  L_{r} D_{d}^{2}  | \cos \gamma |  \nonumber \\
& +  \frac{1}{2} \iint d  \oma  d \oma^{\prime} \rho_{d}( \oma) \rho_{d}( \oma^{\prime})  \frac{\pi}{2} D_{d}^{3}  | \sin \gamma |  
\label{fex}
\end{align}
where $L_{r,d}$ and $D_{r,d}$ denote the length and diameter of the cylindrical building blocks (see \fig{fig:cartoon}(b)).  We assume all polymers and discs to be sufficiently slender, i.e., $L_{r}/D_{r} \gg 1$ and $D_{d}/L_{d} \gg 1$ so that finite-thickness corrections to the excluded volume terms above can be neglected.   Next we formally minimize the free energy with respect to the polymer distribution
\beq
\frac{\delta}{\delta \rho_{r} ( \ell , \oma )} \left ( \frac{ F}{V} - \lambda_{r} \sum_{  \ell} \int d \oma \rho_{r}( \ell, \oma ) \right ) =0
\label{el1}
\eeq
and to the one-body density of the discs
\beq
\frac{\delta}{\delta \rho_{d} ( \oma )} \left ( \frac{ F}{V} - \lambda_{d}  \int d \oma \rho_{d}(  \oma ) \right ) =0
\label{el2}
\eeq
The Lagrange multipliers $\lambda_{r,d}$ ensure that the total concentration of each species (monomers and discs) be preserved.  The coupled Euler-Lagrange (EL) equations can be rendered tractable by expanding the orientation-dependent kernels that depend on the enclosed angle $\gamma$ between the main particle orientation axes, as we will show next.

\section{Molecular weight-distribution from second-polynomial approximation}

A commonly employed method to cast the free energy in a more tractable form is to expand trigonometric functions featuring in the excluded-volume \eq{fex} in terms of a bilinear expansion in Legendre polynomials \cite{lakatos,kayser,lekkerkerker84}. Truncating this expansion after the second-order contribution leads to a simplified theory that has been explored previously for rod-plate mixtures \cite{stroobants1984,jacksonbiax} as well as in the context of  rods with fixed length polydispersity \cite{sollichonsagerP2}. For the present mixture, the approximation should be adequate if the nematic order of either component is not too strong. We  write:
\begin{align}
| \sin \gamma | &= \frac{\pi}{4} - \frac{5 \pi }{32} \pp(\cos \theta) \pp(\cos \theta^{\prime}) + \cdots \nonumber \\
| \cos \gamma | &= \frac{1}{2} + \frac{5 }{8} \pp(\cos \theta) \pp(\cos \theta^{\prime}) + \cdots
\label{p2expansion}
 \end{align}
in terms of the second Legendre polynomials $\pp(x) = \frac{3}{2} x^{2} - \frac{1}{2}$. The  orientation of each particle is described by a polar angle $\theta$ and azimuthal angle $\varphi$ defined with respect to the nematic director $\bn$ (see \fig{fig:cartoon}a and b). Let us define a set of  {\em size-specific} nematic order parameters for the polymer:
\begin{align}
S_{r \ell} = \rho_{r \ell}^{-1} \int d \oma \rho_{r}(\ell, \oma) \pp (\oma \cdot \bn) 
\label{sr} 
\end{align}
with $\rho_{r \ell } = \int d \oma \rho_{r} (\ell, \oma)$ a partial number density of rod segments belonging to polymers of length $\ell L$. Likewise we find for the discs:
\begin{align}
S_{d} = \rho_{d0}^{-1} \int d \oma \rho_{d}( \oma) \pp (\oma \cdot \bn) 
\end{align}
These order parameters allow us to distinguish between an isotropic fluid ($S_{r \ell}=S_{d}=0$), a polymer-dominated uniaxial nematic fluid ($N^{+}$: $S_{r \ell} > 0$, $S_{d} <0$) and a discotic one ($N^{-}$: $S_{r} < 0$, $S_{d} >0$), as sketched in \fig{fig:cartoon}d and e, respectively.  With the aid of these expansions, the excess free energy can be written in terms of a simple bilinear dependence on the nematic order parameter:
\begin{align}
 \frac{F_{ex}}{ V}   \sim & \rho_{r0}^{2}   \left ( 1 - \frac{5}{8}  \bar{S}_{r }^{2} \right ) +  2 q \rho_{r 0} \rho_{d}   \left ( 1 + \frac{5}{4} \bar{S}_{r} S_{d}   \right ) \nonumber \\
& + z \rho_{d}^{2}  \left ( 1 - \frac{5}{8} S_{d}^{2}  \right ) 
\label{fexsim}
 \end{align}
 Here, we have implicitly renormalized the free energy and species densities in terms of the isotropic excluded volume of the monomeric rods  $v_{rr} = \frac{\pi}{4} L_{r}^{2} D_{r}$. The excess free energy thus only depends on the excluded-volume {\em ratios}    $q= v_{rd}/v_{rr}$ and $z =v_{dd}/v_{rr}$  with $v_{rd} = \frac{\pi}{16} L_{r} D_{d}^{2} $ and $v_{dd} = \frac{\pi^{2}}{16} D_{d}^{3}$ denoting the isotropized  monomer-disc and disc-disc excluded volumes, respectively.
Furthermore, the bar denotes a molecular-weight average of the nematic order parameter associated with the polymers:
\begin{align}
\bar{S}_{r} &= \rho_{r0}^{-1}  \sum_{\ell} \rho_{r\ell} S_{r \ell} 
\label{srav}
\end{align}
Similarly, the coupled EL equations may be cast as follows: 
\begin{align}
& \ell^{-1} \ln [ 4 \pi \rho_{r} (\ell, \oma ) \ell^{-1} ]   = \lambda_{r}  + \varepsilon_{b} \ell^{-1} +  a_{r} \pp(\oma \cdot \bn) \nonumber \\ 
&  +\frac{L_{r}}{3\ell_{p}} \frac{\nabla^{2} [ \rho_{r} (\ell , \oma)]^{1/2}}{ [ \rho_{r} (\ell , \oma)]^{1/2} }
\label{lnrhor}
\end{align}
and
\begin{align}
\ln [ 4 \pi  \rho_{d} (\oma )] & = \lambda_{d}  + a_{d} \pp(\oma \cdot \bn)  
\label{lnd}
\end{align}
The uniaxial order parameters that feature in the EL equations are specified as follows:
\begin{align}
a_{r} &=    \frac{5 }{4}  ( \rho_{r0}  \bar{S}_{r} -   2  q\rho_{d0}  S_{d} )  \nonumber \\
a_{d} &=   \frac{5}{4} (  z \rho_{d0}   S_{d}  -  2q  \rho_{r0} \bar{S}_{r} )  
\label{alphabeta}
\end{align}
We  are now equipped to explore the equilibrium polymer length distribution  $\rho_{r \ell} =  \int d \oma \rho_{r} (\ell, \oma)$ corresponding to the basic fluid symmetries we consider (cf. \fig{fig:cartoon}).

\subsection{Isotropic fluid}

In the isotropic phase, all nematic order parameters are strictly zero. Applying conservation of monomers to \eq{lnrhor} and performing some algebraic rearrangements we find a geometric distribution (i.e., the discrete analog of the exponential distribution):
\begin{align}
\rho_{r \ell } & =   \ell e^{ \varepsilon_{b} + \lambda_{r} \ell }  \nonumber \\
&= \ell e^{ \varepsilon_{b}} \left ( 1 - m_{I}^{-1} \right ) ^{\ell} \label{disi}
\end{align}
in terms of the mean aggregation number:
\beq
m_{I} =  \frac{\sum_{\ell} \rho_{a}(\ell) \ell  }{\sum_{\ell} \rho_{a}(\ell) } = \frac{1}{2} \left ( 1+ \sqrt{1+ 4 \rho_{r0} e^{-\varepsilon_{b}} }\right ) 
\label{miso}
\eeq
which, as expected, goes up monotonically with increasing monomer concentration  $\rho_{r0} $ or when the effective temperatures  $\varepsilon_{b}$ grows more negative.  Since there is no global particle alignment whatsoever, the presence of the discs does not influence the polymerization process, and the polymer molecular-weight distribution is independent from the disc concentration.  

\subsection{Uniaxial nematic fluid}

The decoupling of polymeric rods and discs  is no longer  valid  for a nematic fluid where the alignment direction of  of one component is strongly affected by the amount of orientational ``templating" it experiences from the other component.
The polymer density  follows  from  \eq{lnrhor} and can be written in an exponential form:
\begin{align}
4 \pi \rho_{r} (\ell , \oma) &= \ell \exp [\varepsilon_{b} + \ell \lambda_{r} +  \tilde{a}_{r} \ell \pp(t) ] 
\label{rhorfull}
\end{align}
with $t = \cos \theta $. The three basic contributions affecting the polymer molecular weight  distribution  in a (uniaxial) nematic fluid are easily identified in the argument; the first denotes  monomer-monomer bonding while the second term  enforces monomeric mass conservation. The third one is the most interesting one; it encapsulates the templating effect associated with nematization of the discs as per \eq{alphabeta}. Here,  we have introduced $a_{r}$ as a renormalized version of the one in \eq{alphabeta}:
\beq
\tilde{a}_{r} =  a_{r} + \xi 
\eeq
The factor $\xi$ depends on both $a_{r}$ itself and on the polymer persistence length $\ell_{p}$. It accounts for the finite polymer flexibility and vanishes for strictly rigid polymers ($\ell_{p} \rightarrow \infty$). The corresponding expressions are given in the Appendix. As noted previously, the multiplier $\lambda_{r}$ featuring in \eq{rhorfull}  follows from monomer mass conservation:
\beq
  \sum_{\ell=1}^{\infty}  \int d \oma \rho_{r}(\ell, \oma)  = \rho_{r0}
 \label{como}
\eeq
The summation can be resolved analytically and we find:
\beq
\rho_{r0} = e^{\varepsilon_{b}} \frac{1}{2} \int_{-1}^{1} dt \frac{ e^{W(t) }}{ ( e^{W(t)} -1)^{2}}
\eeq
A molecular-weight averaged  nematic order parameter \eq{srav} is then given by:
\beq
\bar{S}_{r} = \rho_{r0}^{-1} e^{\varepsilon_{b}} \frac{1}{2} \int_{-1}^{1} dt  \frac{ \pp(t) e^{ W(t) }}{ (  e^{ W(t)} -1)^{2}}
\eeq
The two conditions above are intricately coupled given that $\tilde{a}_{r}$ depends on both $\bar{S}_{r}$  and $S_{d}$ via \eq{alphabeta}. Convergence of the summation \eq{como}  requires that the argument be negative:
 \beq
 W(t) = \lambda_{r} + \tilde{a}_{r} \pp(t) < 0, \hspace{0.3cm} \textrm{all} \hspace{0.1cm}  t
 \eeq
 Noticing that $ -1/2 \leq|| \pp || \leq 1$ one then finds that $\lambda_{r}$ should satisfy:
 \begin{align}
- \lambda_{r} & <  |\tilde{a}_{r} | \hspace{0.3cm} (N^{+}) \nonumber \\
- \lambda_{r} & <  |\tilde{a}_{r}/2 |\hspace{0.3cm}  (N^{-}) 
 \end{align}
and it is tempting to introduce a rescaled  normalization constant $\lambda_{r}^{\prime}$ that is strictly positive ($\lambda_{r}^{\prime} > 0$) for both phases. With this, we recast:
  \begin{align}
W(t) =  
\begin{cases}
\frac{3}{2} \tilde{a}_{r} (t^{2} -1) - \lambda_{r}^{\prime} \hspace{0.5cm}  (N^{+})   \\
\frac{3}{2} \tilde{a}_{r} t^{2}  - \lambda_{r}^{\prime} \hspace{1.2cm}  (N^{-}) 
\end{cases}
 \end{align}
 Unlike for the isotropic phase, the normalization constant $\lambda_{r}^{\prime}$ can not be resolved in closed form. The  molecular-weight distribution of the polymer follows from integrating \eq{rhorfull} over all orientations $\oma$:
 \beq
 \rho_{r \ell} = \ell e^{\varepsilon_{b}} \frac{1}{2} \int_{-1}^{1} dt e^{\ell W(t)}
 \eeq
The uniaxial nematic order parameter $S_{r \ell}$ associated with a  polymer of length $\ell$ is easily found from:
\beq
S_{r \ell} = -\frac{1}{2} \left \{ 1 + \frac{1}{\tilde{a}_{r} \ell}  - \frac{1}{      F  \left (  \sqrt{3 \tilde{a}_{r} \ell/2}   \right ) \sqrt{    \tilde{a}_{r} \ell /3 }  }  \right \}  
\label{spol}
\eeq
in terms of Dawson's integral $F(x) = e^{-x^{2}} \int_{0}^{x}e^{y^{2}}dy$ \cite{abram}. The discotic nematic order parameter $S_{d}$  easily follows from the above expression upon substituting $\tilde{a}_{r} \ell \rightarrow a_{d}$.   A little reflection of \eq{spol} tells us the following; since $a_{r}$ does not depend explicitly on the aggregation number $\ell$, the nematic order parameter $S_{r \ell}$ must be a monotonically increasing function of the polymerization degree $\ell $; the longer the polymers the stronger their nematic ($ a_{r} >0$) or anti-nematic ($ a_{r} <0$) alignment in the mixed nematic fluid. This effect becomes systematically weaker for increasingly flexible polymers as can easily be inferred from the above expression by comparing  $S_{r \ell}$ versus $\ell$ for rigid polymers $(\xi =0)$ versus the case of slightly flexible ones ($\xi $ nonzero but small) for any given value for $a_{r}$. 

\begin{figure}
  \includegraphics[width=\linewidth]{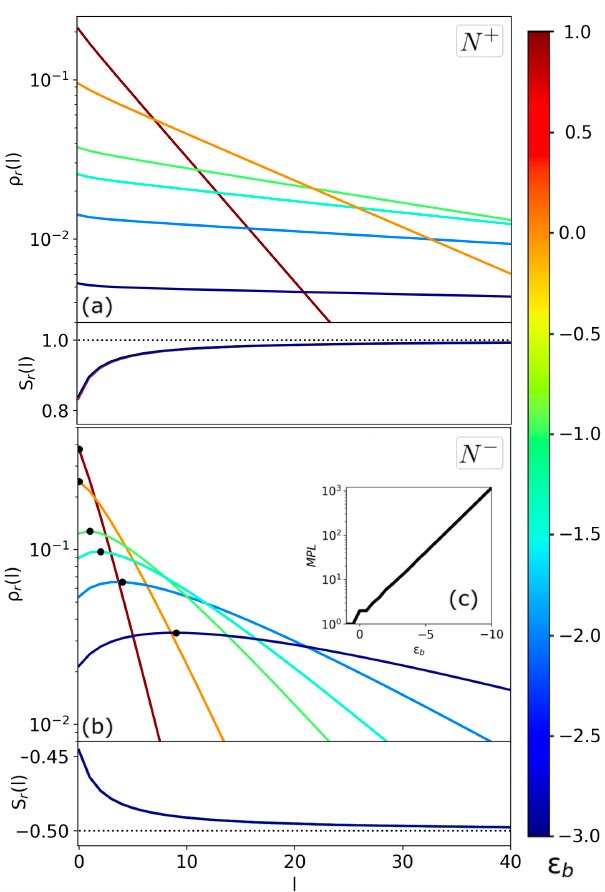}
  \caption{Polymer molecular-weight distributions $\rho_{r \ell}$ and corresponding  uniaxial nematic order parameter $S_{r \ell}$ as a function of the polymer length $\ell$ for (a) a typical polymer nematic $N^{+}$ and (b) discotic nematic phase $N^{-}$. The effective temperature $\varepsilon_{b}$ is color-coded.   
  (c) Most Probable Length (MPL) in terms of the effective temperature $\varepsilon_{b}$. Fixed parameters: persistence length $\ell_{p} = 3$, disc mole fraction $x = 0.5$, overall concentration $\rho = 3$, excluded-volume {\em ratios} $q = \frac{1}{4}\frac{L_{r}}{D_{r}}$ and $z=\pi q$ with monomer aspect ratio $L_{r}/D_{r} =10$.  }
  \label{fig:distributions}
\end{figure}
 
Let us now examine a concrete example by picking a dense uniaxial discotic nematic  doped with polymerizing rods. The polymers are dispersed anti-nematically within the discotic fluid as indicated in \fig{fig:cartoon}(e).   In specifying the shape of the rods and discs, we can distinguish between  so-called {\em symmetric} mixtures \cite{stroobants1984}, in which the excluded-volume  between two monomers, a monomer and a disc and two discs are all equal, so that $q = 1$ and $z = 1$ and {\em asymmetric mixtures} composed of species with  strongly disparate excluded volumes.   Our principal attention goes to the latter systems which arise more naturally in an experimental context when mixing, for instance, tip-associating colloidal rods such as {\em fd} \cite{fraden-tmv-baus,dogic-fraden_fil} with clay platelets \cite{davidson-overview}. The molecular-weight distributions of some mixtures of this nature are shown in \fig{fig:distributions}.

 \fig{fig:distributions}(a) relates to the uniaxial polymer-dominated nematic phase ($N^+$) and demonstrates an exponential probability distribution  whose shape can be tuned by changing the effective temperature of the system. As expected, the tail of the distribution is longer upon decreasing the temperature, which would give longer polymers. A more interesting scenario shows up for the discotic nematic phase ($N^-$) in \fig{fig:distributions}(b), where the distributions are no longer monotonically decreasing. The maximum of the distributions corresponds to the most probable length of the polymers for each system, which depends quite sensitively on the effective temperature as we  observe in \fig{fig:distributions}(c). Reversible polymerization within an anti-nematically organization  thus leads to a strong manifestation of oligomeric polymers at the expense of its monomeric counterparts. We  note that the orientational order associated with the anti-nematic oligomers remains relatively mild (particularly at larger temperature $\varepsilon_{b}$) so that the second-polynomial truncation should not be too severe.

As we will see during the incoming sections of this manuscript, the overall particle concentration and disc molar fraction  associated with \fig{fig:distributions} may correspond to  regions of the phase diagram where the uniform nematic system is in fact thermodynamically unstable with respect to some phase separation. The molecular-weight distributions should therefore be interpreted under the caveat that monophasic nematic fluidity is preserved and that any demixing process is somehow suppressed. We wish to add that the non-monotonic features of the anti-nematic polymer molecular-weight distribution are also present at conditions where monophasic anti-nematic order is found to be stable.  Next we address the  thermodynamic stability of the mixtures  within the context of a Gaussian variational theory. 

\section{Isotropic-nematic phase behavior}

At conditions where (anti-)nematic order is strong, the previously used polynomial-based expansion truncated after ${\mathcal P}_{2}$ [\eq{p2expansion}] is no longer appropriate and  a cumbersome inclusion of multiple higher-order terms becomes necessary \cite{lekkerkerker_coulon84,wensinkbiaxial}. A more technically expedient route towards exploring the thermodynamics of strongly ordered nematic fluids is to use a simple Gaussian parameterization of the orientational probability \cite{odijkoverview,Vroege92}.  Following \cite{wensink_mm2019} we express the polymer molecular-weight distribution in a factorized form:
\beq
\rho_{r} (\ell , \oma ) = \rho_{r \ell} f_{G} (\oma) 
\label{pfac}
\eeq
where $f_{G}$ is a normalized Gaussian distribution with a variational parameter that is proportional to either the amount of nematic  order ($\alpha^{(1)} > 0$) or anti-nematic order  ($\alpha^{(2)} > 0$).  The corresponding Gaussian distributions for the polar angles corresponding to these different nematic symmetries are given by \cite{wensinkrodplate}:
\beq
f_{G} (\oma) \sim  
\begin{cases}
\frac{\alpha^{(1)}}{4 \pi } \exp (-\frac{1}{2}\alpha^{(1)} \theta^{2} ) , \\
\sqrt{ \frac{ \alpha^{(2)} }{ (2 \pi )^{3} } } \exp (-\frac{1}{2} \alpha^{(2)} \psi^{2} )
\end{cases}
\label{gaussians}
\eeq
where $\psi = \frac{\pi}{2} - \theta$ ($-\pi/2 < \psi < \pi/2$) denotes a meridional angle (see \fig{fig:cartoon}a). The Gaussians operates on the domain $0 < \theta < \pi/2$ and must be complemented by its mirror $f_{G} (\pi - \theta )$ for $\pi/2 < \theta < \pi$  given that all nematic phases are required to be strictly apolar. The Gaussian representations are appropriate only for strong nematic order ($\alpha_{r\ell} \gg 1$). They are clearly inadequate for isotropic systems since the probabilities reduce to zero when $\alpha \rightarrow 0$ instead of reaching a constant.  Obviously,  we  apply the same distributions to the discs with $\alpha_{d}^{(1)}$ and $\alpha_{d}^{(2)}$ denoting the variational parameters quantifying the amount of nematic order of the discs. The disc probability density is then equivalent to \eq{pfac}:
\beq
\rho_{d} (\oma ) = \rho_{d0} f_{G} (\oma) 
\eeq
A major advantage of using  Gaussian trial functions is that we may apply  asymptotic expansion of the various free energy contributions \cite{odijkoverview} which are valid in the limit $\alpha \rightarrow \infty$. In particular, it can be shown that the double orientational averages over the sine and cosine in \eq{fex} up to leading order in $\alpha$ take a simple analytic form \cite{wensinkrodplate}.  In the general case in which particles with equal nematic signature (nematic or anti-nematic) do not necessarily have the same degree of alignment the asymptotic averages read:
\begin{align}
\langle \langle | \sin \gamma | \rangle \rangle_{11} & \sim \sqrt{ \frac{\pi}{2} \left ( \frac{1}{\alpha^{(1)}} + \frac{1}{\alpha^{(1\prime)}} \right ) } \nonumber \\  
    \langle \langle | \cos \gamma | \rangle \rangle_{12} & \sim  \sqrt{ \frac{2}{\pi} \left ( \frac{1}{\alpha^{(1)}} + \frac{1}{\alpha^{(2)}} \right ) } \nonumber \\ 
    \langle \langle | \sin \gamma | \rangle \rangle_{22} & \sim \mathcal{F} (\alpha^{(2)}, \alpha^{(2\prime)})
    \label{sinav}
\end{align}
Here,  the double brackets denote the orientational averages featuring in the excess free energy \eq{fex} with $ \langle \cdot \rangle = \int d \oma f_{G} (\oma) $. The symmetry of nematic order clearly matters with the anti-nematic case featuring a distinct logarithmic dependence. The function ${\mathcal F}$ reads in explicit form:
\begin{widetext}
\beq
 {\mathcal F}(\alpha^{(2)},\alpha^{(2 \prime)} ) = \frac{4 \alpha^{(2)} Q -2 (1+Q) \textrm{arctanh} \sqrt{Q}  - (1+Q) \ln (1-Q) +(1+Q) \ln ( 4 \alpha^{(2)} Q) } {2 \pi \alpha^{(2)} Q}
\eeq
\end{widetext}
in terms of the ratio $Q=  \alpha^{(2\prime)}/ \alpha^{(2)}$ with $\alpha^{(2)}$ and $\alpha^{(2\prime)}$ quantifying the anti-nematic order parameters of two polymeric species differing in length. Note that generally, $\alpha^{(2)} \neq \alpha^{(2\prime)}$. The expression becomes a lot more manageable if all polymers are assumed to exhibit an equal degree of alignment, irrespective of their length. Then, $Q=1$ and \cite{wensink2001}:
\beq
{\mathcal F}(\alpha^{(2)})  =  \frac{2}{\pi} \left ( 1 + \frac{\ln \alpha^{(2)}}{2 \alpha^{(2)}} \right )
\eeq
Similar asymptotic expressions may be obtained for the orientational entropy featuring in the ideal free energy \eq{free}. For strong nematic or anti-nematic order we find, respectively \cite{wensinkrodplate}:
\begin{align}
\sigma_{1} = \langle \ln 4 \pi f_{G} (\oma) \rangle_{1} &\sim \ln \alpha^{(1)} -1  \nonumber \\
\sigma_{2} =\langle \ln 4 \pi f_{G} (\oma) \rangle_{2}   &\sim \frac{1}{2} \left ( \ln \alpha^{(2)} + \ln \frac{2}{\pi} -1 \right )  
\label{sigma12}
\end{align}
The worm-like chain entropy \eq{wlc} too can be estimated within in the Gaussian limit which leads to:
\begin{align}
\langle \ln f_{G}^{1/2} (\oma) \nabla^{2} f_{G}^{1/2} (\oma)  \rangle_{1} & \sim -\frac{\alpha^{(1)}}{2} \nonumber \\ 
\langle \ln f_{G}^{1/2} (\oma) \nabla^{2}  f_{G}^{1/2} (\oma)  \rangle_{2} & \sim  
-\frac{\alpha^{(2)}}{4} 
\end{align}
We infer that the loss of conformational entropy of an {\em anti-nematic} polymer is half that of a nematic polymer. This suggests that a worm-like chain is able to retain more of its internal configurations when aligned anti-nematically than in a nematic organization of equal strength. With all the orientational averages specified, we  now turn to computing the free energy and its derivatives.

\subsection{Polymer nematic phase ($N^{+}$)}

We now focus on the case of the polymer-dominated nematic phase which is expected to be stable at elevated monomer concentration and low disc mole fraction.  Inserting the asymptotic orientational averages formulated above into the corresponding entropic contributions in \eq{free} we obtain the following algebraic expression for free energy density (in units thermal energy $k_{B}T$ per randomized monomer excluded volume $v_{rr}$):
\begin{align}
   & \frac{F^{N^{+}}}{V}  \sim  \sum_{\ell }  \rho_{r \ell} \ell^{-1}  \left [ \ln \rho_{r \ell} \ell^{-1}  -1 - \varepsilon_{b}  + \sigma_{1}(\alpha_{r \ell}) \right ] \nonumber \\
   & + \rho_{d0} \left [ \ln \rho_{d0} -1 + \sigma_{2}(\alpha_{d}) \right ]  \nonumber \\ 
    & + \frac{1}{3\ell_{p}} \sum_{\ell } \rho_{r \ell} \alpha_{r \ell} + \sum_{\ell,\ell^{\prime}} \rho_{r \ell} \rho_{r \ell^{\prime}} h_{r\ell r\ellp} \nonumber \\
    & + 2 q \rho_{d0} \sum_{\ell} \rho_{r \ell} h_{r\ell d} + z \rho_{d0}^{2} \frac{8}{\pi^{2}} \left ( 1+ \frac{\ln \alpha_{d}}{2 \alpha_{d}}   \right ) 
\end{align}
with $h_{ij}$ is short-hand notation for:
\beq
h_{ij} = \sqrt{ \frac{8}{\pi}   \left ( \frac{1}{\alpha_{i}} + \frac{1}{\alpha_{j}} \right )} 
\eeq
where $\alpha_{i}$ and $\alpha_{j}$ should be considered dummy variables for the species-dependent nematic order parameters as specified by the indices $i$ and $j$. 
For later reference we also define:
\beq
g_{ij} = \left (  \frac{8}{\pi} \right)^{1/2}  \left ( 1+ \frac{\alpha_{i} }{\alpha_{j }} \right )^{-1/2}
\eeq
At equilibrium, the species-dependent nematic order parameters $\alpha_{r\ell}$ and $\alpha_{d}$ follow from the minimum conditions:
\begin{align}
    \frac{\partial F/V}{\partial \alpha_{r \ell,d}} &=0 
    \label{dalpha}
\end{align}
The expressions above can be simplified considerably by noting that a small amount  of backbone flexibility causes the nematic alignment to fully decorrelate from the polymer contour length. We then approximate $\alpha_{r \ell} \approx \alpha_{r \ell^{\prime}} = \alpha_{r}$, independent from $\ell$.   Applying \eq{dalpha} we obtain a set of simple algebraic equations:
\begin{align}
      m_{N^{+}}^{-1} \alpha_{r}^{1/2}  &= - \frac{1}{3\ell_{p}} \alpha_{r}^{3/2} + \frac{2}{\sqrt{\pi}}  \rho_{r0} + q \rho_{d0}  g_{rd} \nonumber \\
         \alpha_{d}^{1/2} &=  z \rho_{d0} \frac{8}{\pi^{2}}\left (  \frac{ \ln \alpha_{d} -1}{\alpha_{d}^{1/2}} \right ) 
  + 4 q \rho_{r0} g_{dr} 
  \label{alphaplus}
\end{align}
with $m_{N^{+}}$ the mean aggregation number in the polymer nematic phase.
The  molecular-weight distribution  now becomes strictly exponential, as for the isotropic phase. We write:
\begin{align}
\rho_{r \ell } & = \ell e^{ \tilde{\varepsilon}_{b}} \left ( 1 - m_{N^{+}}^{-1} \right ) ^{\ell} \label{disn}
\end{align}
with an {\em effective} potential $\tilde{\varepsilon}_{b}$ that depends on the orientational entropy:
\beq
\tilde{\varepsilon}_{b} = \varepsilon_{b} - \sigma_{1}(\alpha_{r}) 
\label{tempnplus}
\eeq
Given that $\sigma_{1} >0$, the effective temperature is {\em lower} than the bare one, so that polymerization in the nematic phase is stronger than in the isotropic fluid, as is well established \cite{vdschoot1994epl,vdschoot1994la}. The mean aggregation number in the nematic phase  has an analogous form to \eq{miso}):
\beq
m_{N^{+}}  = \frac{1}{2} \left ( 1+ \sqrt{1+ 4 \rho_{r0} e^{-\tilde{\varepsilon}_{b}} }\right ) 
\eeq
The chemical potentials are obtained from the standard thermodynamic relations $\mu_{r,d}  = \partial (F/V) /\partial \rho_{r0,d0}$. The contribution from the polymers reads:
\begin{align}
\mu_{r}^{N^{+}}  \sim & \ln ( 1 - m_{N^{+}}^{-1}) + \frac{1}{3 \ell_{p}}  +  m_{N^{+}}^{-1} \sigma_{1}(\alpha_{r})  \nonumber \\
& + 2  \rho_{r0}   \frac{4}{\sqrt{\pi \alpha_{r}}}  +   2 q \rho_{d0} h_{rd}
+ \varepsilon_{b} 
\end{align}
while for the discs we find:
\begin{align}
\mu_{d}^{N^{+}}  \sim  & \ln \rho_{d0}  + \sigma_{2}(\alpha_{d})  +   2 q \rho_{r0}  
h_{rd} + 2 z \rho_{d0} 
\frac{8}{\pi^{2}} \left ( 1+ \frac{\ln \alpha_{d}}{2 \alpha_{d}} \right )
\end{align}
The osmotic pressure follows from  the thermodynamic relation $-P = (F - N \mu)/V$ leading to:
\begin{align}
P^{N^{+}} \sim & e^{\tilde{\varepsilon}_{b}} (m_{N^{+}} -1) 
+ \rho_{d0}  +  \rho_{r0}^{2}  \frac{4}{\sqrt{\pi \alpha_{r}}}
 \nonumber \\ 
& +   2 q \rho_{r0}  \rho_{d0} h_{rd}
 + z \rho_{d0}^{2} 
  \frac{8}{\pi^{2}} \left ( 1+ \frac{\ln \alpha_{d}}{2 \alpha_{d}}   \right )
\label{pressurenp}
\end{align}
Note that all pressures are implicitly renormalized in units of thermal energy $k_{B}T$ per monomer excluded volume  $v_{rr}$.

\subsection{Discotic nematic phase ($N^{-}$)}

Repeating the previous steps for the discotic nematic through simple bookkeeping we write for the free energy of the discotic phase:
\begin{align}
   & \frac{F^{N^{-}}}{V}  \sim  \sum_{\ell }  \rho_{r \ell} \ell^{-1}  \left [ \ln \rho_{r \ell} \ell^{-1}  -1 - \varepsilon_{b}  +\sigma_{2} (\alpha_{r \ell})  \right ] \nonumber \\
   & + \rho_{d0} [ \ln \rho_{d0} -1 + \sigma_{1}(\alpha_{d})  ] + \frac{1}{6\ell_{p}} \sum_{\ell } \rho_{r \ell} \alpha_{r \ell} \nonumber \\ 
   & + \sum_{\ell,\ell^{\prime}} \rho_{r \ell} \rho_{r \ell^{\prime}} 
    \frac{4}{\pi} {\mathcal F} (\alpha_{r \ell}, \alpha_{r \ell^{\prime}}) \nonumber \\
    & + 2 q \rho_{d0} \sum_{\ell} \rho_{r \ell}  h_{r\ell d} + z \rho_{d0}^{2} \frac{4} {\sqrt{ \pi \alpha_{d}}}
\end{align}
The corresponding minimum conditions for the variational parameters under the assumption that all polymer species experience the same degree of orientational order ($\alpha_{r \ell} = \alpha_{r \ell \prime} = \alpha_{r}$) are as follows: 
\begin{align}
     \frac{1}{2} m_{N^{-}}^{-1}  \alpha_{r }^{1/2} &=  - \frac{1}{6\ell_{p}} \alpha_{r }^{3/2} +  \rho_{r0}  \frac{8}{\pi^{2}} 
    \left ( \frac{\ln  \alpha_{r} - 1}{\alpha_{r}^{1/2}} \right )   + q \rho_{d0}  g_{rd} \nonumber \\
 \alpha_{d}^{1/2} &=  z \rho_{d0} \frac{2}{\pi^{1/2}}  + 2 q \rho_{r0} g_{dr}
 \label{alphamin}
\end{align}
The molecular-weight distribution is  analogous to \eq{disn} but with the effective temperature now reading:
\beq
\tilde{\varepsilon}_{b} = \varepsilon_{b} - \sigma_{2}(\alpha_{r}) 
\label{tempnmin}
\eeq
which, as for the case of the polymer nematic phase suggests that particle alignment facilitates polymer growth, although less so for anti-nematic polymers since generally $\sigma_{2}  < \sigma_{1}$ (\eq{sigma12}).  The chemical potential of the polymers and the discs are given by, respectively:
\begin{align}
 \mu_{r}^{N^{-}}   \sim  
& \ln ( 1 - m_{N^{-}}^{-1}) + \frac{1}{6 \ell_{p}}  +  m_{N^{-}}^{-1} \sigma_{2}(\alpha_{r})  \nonumber \\
& + 2  \rho_{r0} 
\frac{8}{\pi^{2}} \left ( 1+ \frac{\ln \alpha_{r}}{2 \alpha_{r}} \right )
+   2 q \rho_{d0}  h_{rd}
+ \varepsilon_{b} \nonumber \\
 \mu_{d}^{N^{-}}  \sim  & \ln \rho_{d0}  + \sigma_{1}(\alpha_{d})  +   2 q \rho_{r0}  
h_{rd} + 2 z \rho_{d0}  \frac{4} {\sqrt{ \pi \alpha_{d}}}
\end{align}
Finally, the pressure of the $N^{-}$ phase reads:
\begin{align}
P^{N^{-}} \sim & e^{\tilde{\varepsilon}_{b}} (m_{N^{-}} -1) + \rho_{d0} +  \rho_{r0}^{2}  \frac{8}{\pi^{2}} \left ( 1+ \frac{\ln \alpha_{r}}{2 \alpha_{r}}   \right )
 \nonumber \\ 
& +   2 q \rho_{r0}  \rho_{d0}  h_{rd}
 + z \rho_{d0}^{2}  \frac{4}{\sqrt{\pi \alpha_{d}}}
\label{pressurenm} 
\end{align}
The thermodynamics of the isotropic phase is easily established from the original free energy \eq{free} because the randomized excluded volumes becomes simple constants, namely  $\langle \langle | \sin \gamma | \rangle \rangle = \pi/4$ and $\langle \langle  | \cos \gamma | \rangle \rangle  = 1/2$. We thus obtain the following expressions for the chemical potentials in the isotropic fluid \cite{wensink_mm2019}:
\begin{align}
  \mu_{r}^{\textrm I} & \sim  \ln ( 1- m_{I}^{-1} )  + 2  \rho_{r0}  +   2 q \rho_{d0}  + \varepsilon_{b} \nonumber \\
   \mu_{d}^{\textrm I} & \sim \ln \rho_{d0}  +   2 z \rho_{d0} + 2 q \rho_{r0} 
  \end{align}
The osmotic pressure  combines the ideal gas and excluded volume contributions and reads:
\begin{align}
   P^{\textrm I}  &\sim  e^{\varepsilon_{b}} (m_{I} -1) + \rho_{d0}  +  \rho_{r0}^{2}
+   2 q \rho_{r0}  \rho_{d0}  + z \rho_{d0}^{2} 
\end{align}
Binodals denoting coexistence between phases of any symmetry may be established from equating chemical potentials and pressures in conjunction with the minimum conditions for the nematic variational parameters, where relevant.  Phase diagrams can be represented in a pressure-composition  ($P-x$) plane or, alternatively, in a density-density representation using $\rho_{r0} = c (1-x)$ and $\rho_{d0}  = cx$ in terms of the overall particle concentration $c$ and disc mole fraction ($0<x<1$). In order to remain consistent with the Gaussian approximation adopted in our analysis, we will focus on {\em asymmetric} mixtures characterized by both monomer-disc and disc-disc excluded volumes being much larger than the monomer-monomer one. The considerable excluded-volume disparity thus ensures that the nematic order of all components be sufficiently strong. Concretely, we impose that $\alpha_{r,d} >5$  for all numerical results to be self-consistent.

\section{Phase diagrams}

\begin{figure*}[ht]
  \includegraphics[width=0.9\textwidth]{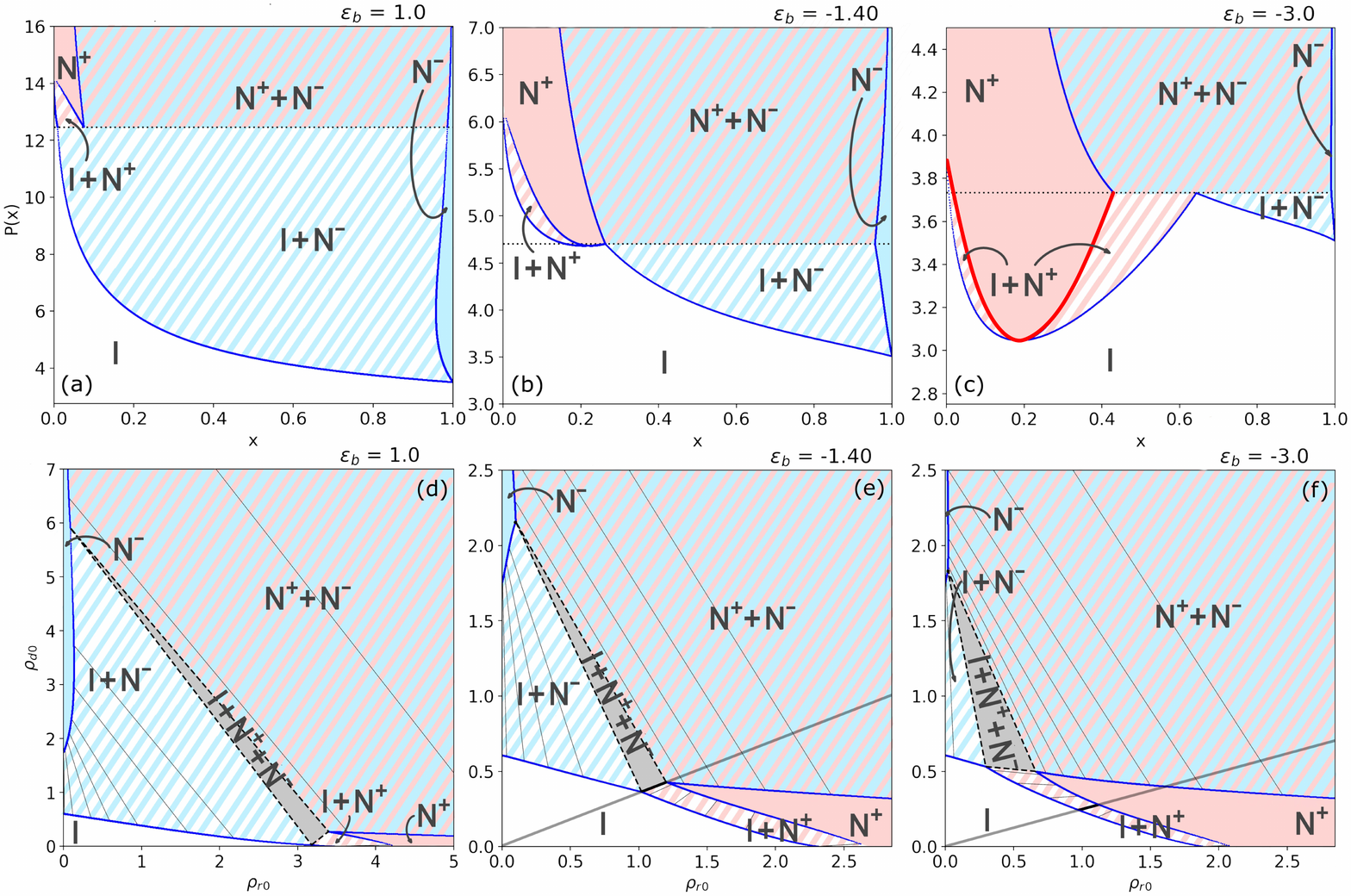}
  \caption{Overview of the isotropic $I$ (white) -  polymer nematic $N^+$ (red) - discotic nematic $N^-$ (blue) phase diagrams for a mixture of discs and reversibly polymerizing weakly flexible rods at various effective temperatures $\varepsilon_{b}$. Two types of phase diagrams are represented: osmotic pressure $P$ versus disc mole fraction $x$ (top panels) and concentration of discs $\rho_{d0}$ versus concentration of rods $\rho_{r0}$ (bottom panels). Fixed parameters: persistence length $\ell_{p} = 3$, excluded-volume {\em ratios} $q = \frac{1}{4}\frac{L_{r}}{D_{r}}$ and $z=\pi q$ where $L_{r}/D_{r} =10$. The presence of a negative {\em azeotrope} is indicated in panels (e) and (f) as a bold black line, which is shown to be parallel to the dilution line (grey diagonal shown in (e) and (f)).}
  \label{fig:L10}
\end{figure*}

\fig{fig:L10} presents an overview of the isotropic-nematic phase diagram for a mixture of reversibly polymerizing rods and discs at three different temperatures.  The choice of excluded-volume parameter $q$ and $z$ is inspired by the typical dimensions of experimentally realizable anisotropic colloids, where the  monomeric rods and discs usually have {\em equal} largest dimensions ($L_{r} = D_{d}$). The monomer aspect ratio $L_{r}/D_{r}$ can be chosen freely but we fix it here at  $L_{r}/D_{r} = 10$. The disc aspect ratio is not constrained as long as the discs are sufficiently thin ($D_{d}/L_{d} \gg 1$). In this study we keep the persistence length fixed at $\ell_{p} =3$. We found that variations up to $\ell_{p} =10$ (corresponding to stiffer monomers) did not lead to major changes in the phase behavior.  For practical reasons we refrained from exploring the near-rigid rod limit ($\ell_{p} \rightarrow \infty$) which is known to cause the polymers to grow to unphysically large lengths \cite{vdschoot1994la}.

Several key trends in the phase diagrams can be discerned. First of all, \fig{fig:L10}(a) correspond to high-temperature scenario in which reversible polymerization happens on very limited scale. The shape-dissimilar nature of the mixture translates into  diagram that is highly asymmetric about the equimolar point $x=0.5$. Second, demixing is  prominent given the large range of monomer-disc compositions where the mixture fractionates into strongly segregated uniaxial nematic phases [\fig{fig:L10}(a)]. Only at very low osmotic pressures, where particle exclusion effects are relatively weak, does the mixture remain miscible throughout the entire composition range. We further observe that the discotic nematic $N^{-}$ can be stabilized over a relatively broad pressure range, while the polymer nematic ($N^{+}$)  only features at elevated pressures, where polymerization is strong enough for the long polymers to align into a conventional nematic organization with the discs interspersed anti-nematically. The phase diagram also features a triple $I-N^{+}-N^{-}$ equilibrium in agreement with previous predictions  \cite{galindo2,wensinkrodplate} and experiment \cite{kooijlangmuir2000,woolston2015} for discs mixed with {\em non-}polymerizing rods. 

\begin{figure}
  \includegraphics[width= .8\linewidth]{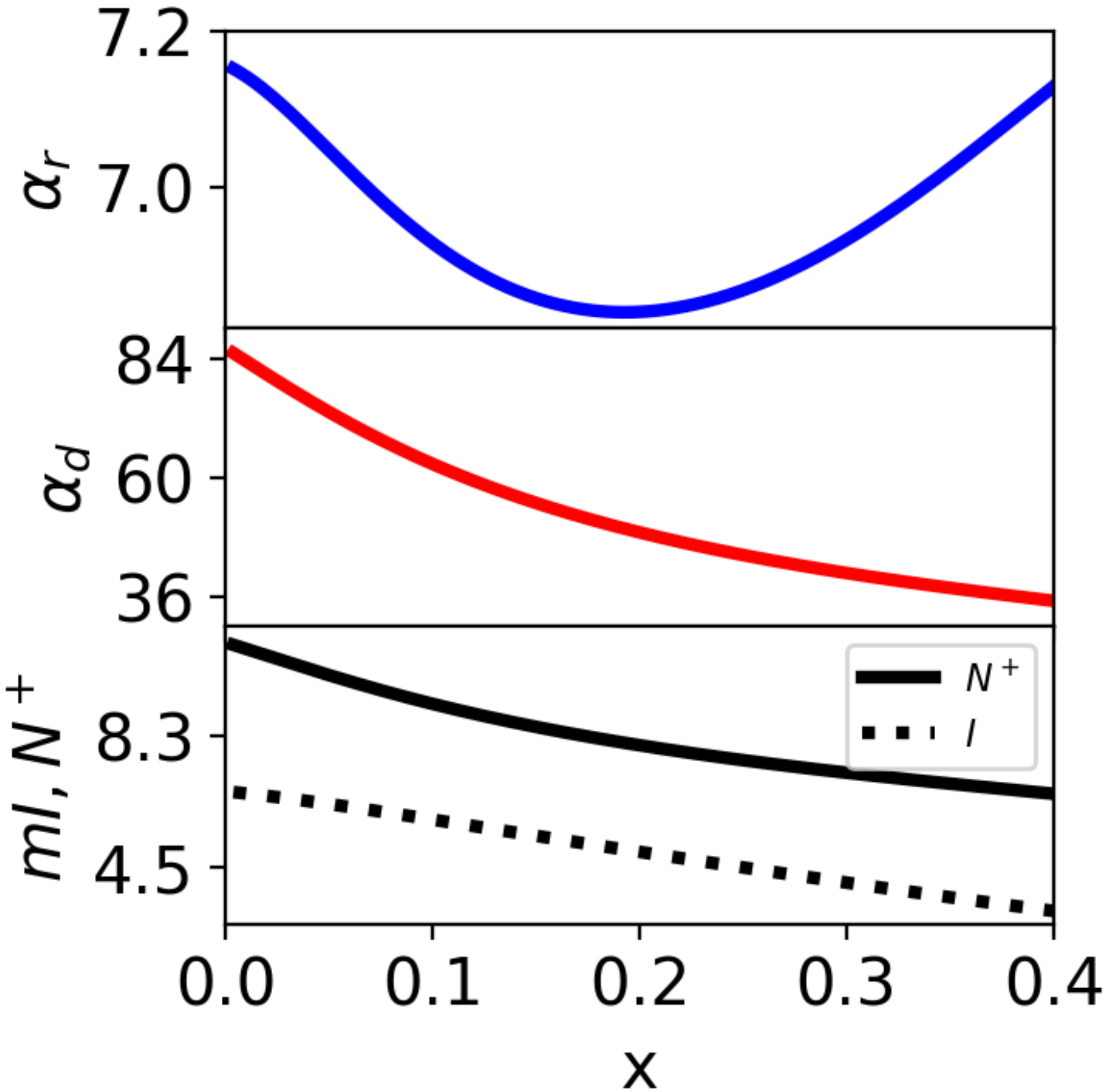}
\caption{Nematic order of the polymers ($\alpha_{r}$) and discs ($\alpha_{d}$) and polymer mean aggregation number ($m$) of  the nematic $N^{+}$ phase in coexistence with the isotropic phase $I$  across the azeotropic region. The corresponding binodal in \fig{fig:L10}(c) has been indicated in red. } 
  \label{fig:azeotrope}
\end{figure}

Reducing the temperature stimulates polymer growth and, consequently, enhances the stability window for the polymer-dominated nematic  [\fig{fig:L10}(b) and (c)]. Reversible polymerization thus renders the phase diagrams less asymmetric.  At the same time, the osmotic pressure (and concomitantly the particle concentrations) at which nematic order occurs drops significantly as polymerization becomes more prominent. Furthermore, the $I-N^{+}$ binodals develop a remarkable (negative) {\em azeotrope} which in \fig{fig:L10}(b) coincides with the triple pressure. Under these conditions,  coexistence occurs between a discotic nematic,  a polymer nematic and an isotropic fluid with the latter two having the same monomer-disc composition. At lower temperature the azeotrope comes out more prominently at $x\approx 0.2$ (\fig{fig:L10}(c)). In the density density representations shown in the bottom panels, the azeotrope manifests itself at the point where the tie line connecting the monomer and discs concentrations of the coexisting $I$ and $N^{+}$ phases coincides with the dilution line. The latter are straight lines emanating from the origin along which the overall particle concentration changes but the monomer-disc composition is preserved.  It can  be gleaned that upon following a dilution line at, for instance, $x=0.2$  the sequence of phase transitions encountered depends strongly on temperature. At high temperature [\fig{fig:L10}(a)] the isotropic fluid first transforms into $N^{-}$, then develops a triphasic $I-N^{+}-N^{-}$ equilibrium. At low temperature, however, a polymer nematic is formed first, followed by a binematic $N^{+}-N^{-}$ coexistence while the triphasic equilibrium does not show up at all unless the monomer concentration is significantly increased.   \fig{fig:azeotrope} provides insight into the change of nematic order of the polymers and discs as well as the mean aggregation number of the $N^{+}$ across the azeotrope. In view of their considerable excluded volume the discs are way more ordered than the polymers ($\alpha_{d} >\alpha_{r}$).  Increasing the  mole fraction of discs  reduces the nematic order of both components, though the decrease is much more significant for the discs than the change of $\alpha_{r}$ for the polymers which in fact develops a minimum at the azeotrope.

\begin{figure*}[ht]
  \includegraphics[width= \linewidth]{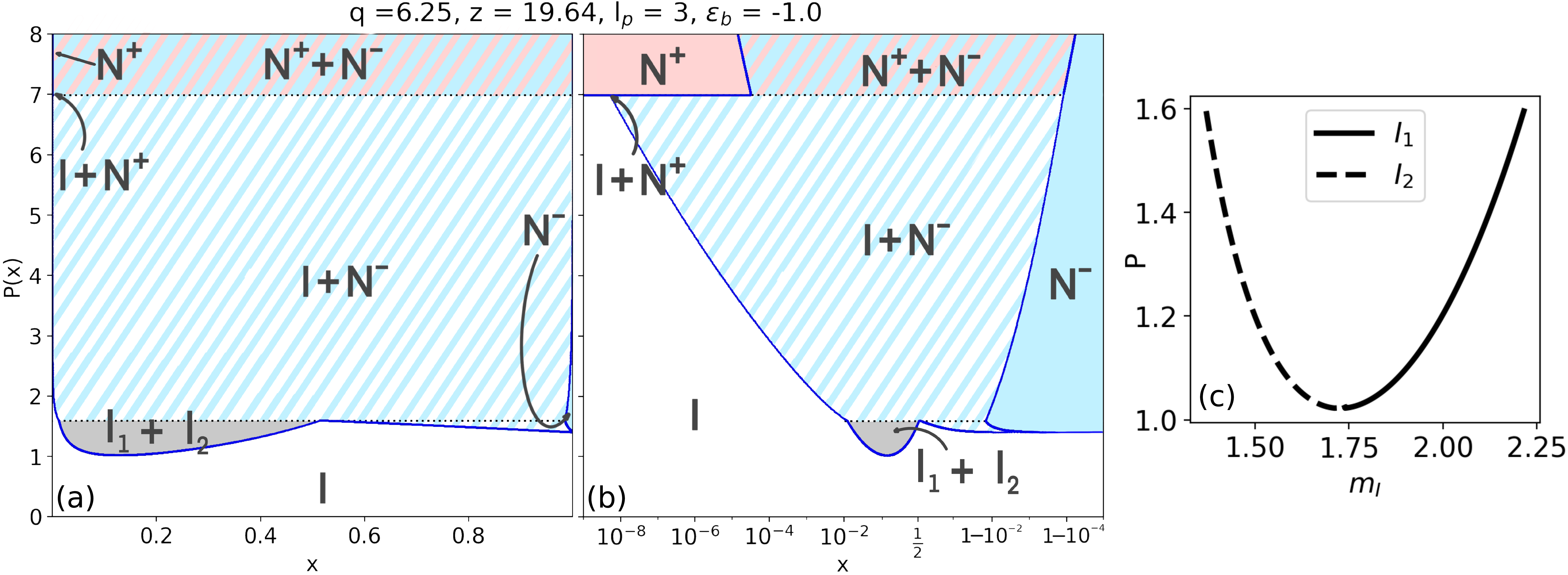}
\caption{Phase diagram in the osmotic pressure-composition ($P-x$) representation with the following parameters: persistence length $\ell_{p} = 3$, effective temperature $\varepsilon_{b} = -1$, excluded-volume {\em ratios} $q = \frac{1}{4}\frac{L_{r}}{D_{r}}$ and $z=\pi q$ corresponding to a monomer aspect ratio $L_{r}/D_{r} =25$. The disc mole fraction $x$ is plotted on a linear scale (a) and on a logarithmic scale (b) to highlight the behavior close to single-component systems (pure polymers $x=0$, and pure discs $x=1$). Note the presence of a coexistence between an isotropic gas and fluid  phase ($I_1$ and $I_2$) with different discs compositions (grey region). (c) Comparison of mean aggregation numbers $m_{I}$ between $I_{1}$ and $I_{2}$ for a given pressure. $I_{1}$ corresponds to the phase at the lowest disc mole fraction $x$. }
  \label{fig:coexistence}
\end{figure*}

We move on to explore a similar mixture featuring more slender rod monomers, namely $L_{r}/D_{r} = 25$. The resulting phase diagram is shown in \fig{fig:coexistence}. The asymmetry of the mixture is now very strong with the monophasic  $N^{+}$ and $N^{-}$ regions being largely unstable except for strongly purified systems  ($x$ close to $0$ or $1$)  [\fig{fig:coexistence}(b)]. Qualitatively, the phase diagram resembles the one in \fig{fig:L10}(a), but the isotropic fluid  undergoes a gas-liquid-type phase separation producing two phases differing in composition. The $I_{1}$-phase may be associated with a discotic colloidal gas, and  $I_{2}$  with its liquid counterpart.   The demixing is driven by the extreme excluded-volume difference between the rod monomers and the discs. This phenomenon has been reported for (non-polymerizing) rod-disc mixtures in Ref. \cite{varga2002}, where the effect was ascribed to a {\em depletion} of discs by the much smaller rods. Isotropic-isotropic demixing has  been more generally observed when mixing different shapes dominated by hard-core repulsion \cite{dijkstra1994}, including thin and thick rods \cite{vanRoij97}, spheres and discs   \cite{chen2015,aliabadi2016} and discs  differing in diameter \cite{phillips_pre2010}. It has also been observed in thermotropic LC-solvent mixtures where the effect is primarily of enthalpic origin and is caused by specific interactions between the LC forming molecules and the solvent \cite{matsuyama1996,reyes2019}.  It is well known that mixing colloids with non-adsorbing polymer depletants creates an effective attraction between the colloids which is entirely of entropic origin and may drive various types of demixing mechanisms \cite{LekkerkerkerTuinier2011}. In our case, the depletion effect is however less clear-cut given that the ``depletants"   reversibly polymerize into a wide array of different sizes \cite{peters2021} and experience  orientation-dependent volume-exclusion interactions which are usually ignored in colloid-polymer models. Moreover the average polymer size depends, via \eq{miso}, on the monomer concentration which is different in the gas and liquid phases. \fig{fig:coexistence}(c) demonstrates that the difference in mean aggregation number between the two isotropic phases is in fact quite small, with the disc-rich fraction harboring slightly longer polymers.   Note that the presence of isotropic-isotropic demixing gives rise to a low-pressure triple equilibrium where both phases coexist with a discotic nematic $N^{-}$.  

\section{Quadruple fluid coexistence}

\begin{figure*}[ht]
  \includegraphics[width=0.8 \linewidth]{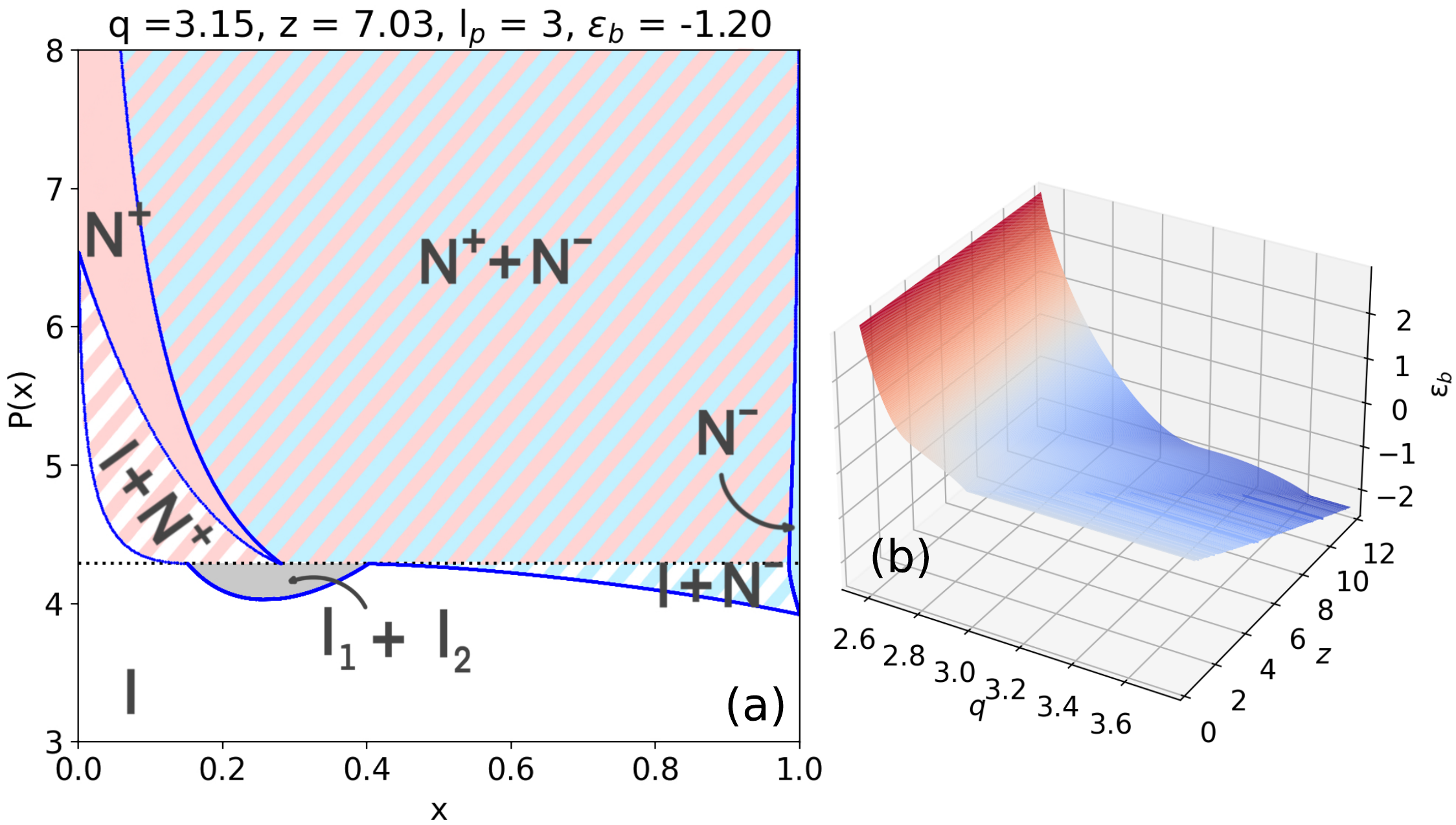}
\caption{(a) Phase diagram in the osmotic pressure-composition ($P-x$) representation showing a $I_{1}$-$I_{2}$-$N^{+}$-$N^{-}$ quadruple point at $P = 4.29$. For this particular case, $\ell_{p} = 3$, $\varepsilon_{b} = -1.2$, $L_{r}/D_{r} =25$ and $D_{d}/L_{r} = 0.7$. (b) Visualization of combinations of rod-disc excluded-volume ratio ($q$ and $z$) and temperature $\varepsilon_{b}$ where  a $I_{1}-I_{2}-N^{+}-N^{-}$ quadruple coexistence is possible.  }
\label{fig:qp}
\end{figure*}

At this stage, one might wonder whether a mixtures could be designed in which  the two separate triple equilibria  in \fig{fig:coexistence} were to join into a  {\em quadruple} coexistence featuring all fluid phases. In \fig{fig:qp}(a) we demonstrate that this scenario is indeed possible. For the particular mixture shown there, the rod monomers and discs no longer have equal largest dimensions ($L_{r}= D_{d}$) but the disc diameter is somewhat smaller than the rod length, namely $D_{d} =0.7 L_{r}$ while the rods are kept sufficiently slender ($L_{r}/D_{r} = 25$). The  excluded-volume asymmetry is then sufficiently reduced to make the two triple points coincide and generate a simultaneous coexistence between two isotropic and two nematic phases, each differing in monomer-disc composition and overall particle concentration. This mixture is by no means  unique and belongs to a family of monomer-disc size ratios where a remarkable $I_{1}$-$I_{2}$-$N^{+}$-$N^{-}$ quadruple point could be encountered, as illustrated by the colored manifold in \fig{fig:qp}(b). This result provides important guidance if one wishes to explore these intricate multi-phase equilibria in real-life mixtures featuring reversibly polymerizing rods mixed with colloidal platelets.

 At this point we wish to draw a connection with recent theoretical explorations of polymer depletion on purely monomeric colloidal rods which have revealed similar multi-phase equilibria involving one-dimensional periodic smectic structures as well as fully crystal states \cite{peters_prl2020}. Similar phenomena involving isotropic-nematic-columnar quadruple points had been reported previously for disc-polymer mixtures \cite{gonzalez2017}. In those studies, the multiphase equilibria emerge from an effective one-component theory based on free-volume theory where   polymeric depletants, envisaged as fixed-shape spherical particles that do not interact with one another, are depleted from the surface of the colloidal rod due to volume exclusion as per the original Asakura-Oosawa model \cite{ASAKURA54,ASAKURA58,Vrijdepletie}. In our work, the depletion effect is strongly convoluted since all components (polymer species and discs alike) are explicitly correlated, albeit on the simplified second-virial level. Furthermore, high-density crystal phases with long-ranged positional order are not considered in the present study since their stability requires strong uniformity in particle shape \cite{mederos_overview2014}, which is not the case in our mixtures. In fact, even for basic mixtures of non-associating hard rods mixed with hard discs the full phase behavior at conditions of elevated particle packing  remains largely elusive to this day. Large-scale  numerical simulations or density-functional computations are needed  to overcome the limitations of the simple second-virial approach taken here, but these are technically challenging to implement for dense multi-component systems. 
 
 The results gathered in \fig{fig:qp}  illustrates the possibility of generating four different fluid textures emerging from reversibly changing excluded-volume-driven interactions alone, without the need to invoke attractive interparticle forces. This could bear some relevance on the emergence of functionality through liquid-liquid type phase separation in biological cells which are composed of biomolecules possessing a multitude of different shapes, some of them  controlled by reversible association \cite{hyman2014,shin2017}.

\section{Conclusions}

We have explored the phase behaviour of a simple model for thermoresponsive supramolecular rods mixed with discotic particles. Possessing attractive tips the rod monomers reversibly associate into polymers that retain their basic slender rod shape and experience only a limited degree of backbone flexibility. The interaction between the species is assumed to be  of steric origin such that basic shape differences between the constituents, more specifically the excluded-volume disparity, plays a key role in determining the prevailing liquid crystal symmetry. The principal ones are a   polymer nematic ($N^{+}$) composed of nematic polymer interspersed with an anti-nematic organization of discs and a discotic nematic  ($N^{-}$) in which the polymers are dispersed anti-nematically.   Lowering   temperature stimulates  polymer growth which  enlarges the stability window for the $N^{+}$ phase.  The phase diagram  develops a marked {\em azeotrope} upon increasing the mole fraction of added discs which indicated that the  polymer nematic is stabilized by the addition of non-adsorbing rigid discs provided their mole fraction remains small. 
The polymer-dominated nematic phase eventually becomes destabilized at larger mole fractions where mutual disc alignment disrupts  the nematic  order of the polymers in favour of the formation of a discotic nematic phase in which the polymers self-organize into an anti-nematic structure. The corresponding molecular weight distribution functions strongly deviates from the usual exponential form and becomes non-monotonic with a maximum probability associated with oligomeric aggregates. Enhancing the shape-asymmetry between the rod monomers and discs induces a depletion-driven demixing of the isotropic fluid and opens up the possibility of a quadruple existence featuring two isotropic phase along with the fractionated polymer and discotic nematic phases. Such quadruple points could be expected in wide range of mixed-shape nematics involving supramolecular rods templated by discs and highlight the possibility of multiple liquid symmetries (both isotropic and anisotropic) coexisting in  mixtures of anisotropic colloids with reversible and thermoresponsive shape-asymmetry without cohesive interparticle forces.   Future explorations should aim at a more careful assessment of biaxial nematic order, ignored in the present study,  which could develop in near-equimolar rod-disc mixtures provided they are stable against global demixing (see Appendix B for tentative discussion). Polymerizing rods and discs with finite particle thickness and low shape asymmetry may favor the emergence of liquid crystals possessing lamellar, columnar or fully crystalline signatures  \cite{peroukidis2010} which may be addressed using  computer simulation models along the lines of Refs. \cite{kuriabova2010,nguyen2014,perouklapp2020}. Inspiration for such mixed-shape  lamellar structures  could be drawn from bio-inspired supramolecular liquid crystals \cite{safinya2013} such as, for example, the `sliding columnar phase' and similar stacked architectures  observed in cationic lyposome-DNA complexes \cite{wong2000,ohern1998} which are essentially made up of mixed planar and rod-shaped architectures.

\begin{acknowledgments}

We acknowledge financial support  from the French National Research Agency (ANR) under  grant ANR-19-CE30-0024
``ViroLego".

\end{acknowledgments}

\section*{Conflict of interest}

The authors have no conflicts to disclose

\section*{Appendix A: Renormalized ${\mathcal P}_{2}$ approximation for slightly flexible polymers}

We seek a simple perturbation theory for the one-body density   \eq{lnrhor}  of near-rigid polymers characterized by a finite persistence length $\ell_{p}$. Let us attempt the following generalization of the probability density distribution for the polymers: 
\beq
 \rho_{r} (\ell , \oma) =  \ell e^{\varepsilon_{b} + \lambda_{r} \ell} e^{ \ell (a_{r}+ \xi) \pp(\oma) }
\eeq
with $\xi$ representing a correction induced by the {\em internal} orientational entropy of the polymer due to a small degree of worm-like chain flexibility. 
Inserting this expression into the worm-like chain contribution (last term) in the EL equation  \eq{lnrhor},  substituting $\nabla^{2} = \partial_{t} (1-t^{2}) \partial_{t}$ and $t= \cos \theta$,  we find that for the uniaxial symmetry:
\begin{align}
\frac{\nabla^{2}  \rho_{r}^{1/2}}{  \rho_{r} ^{1/2} }  = & \frac{3}{4} \tilde{a}_{r}^{2} + \left  ( \frac{3}{2}  \tilde{a}_{r} ^{2}  - 3 \tilde{a}_{r} \right ) \pp(t) + {\mathcal O}(t^{4})
\label{roro1}
\end{align}
where $\tilde{a}_{r} = a_{r}  + \xi$ denotes a rescaled alignment amplitude for the polymer. 

\subsection*{Anti-nematic polymers}
We expect that neglecting the fourth-order term will be fairly harmless in a strongly anti-nematic state  where $t $ is generally very small (since $\theta \sim \pi/2 $ for most polymers). This situation is naturally encountered in the  $N^{-}$ phase where $a_{r} \ell \ll 0 $ in particular for the long polymers.  The constant  in \eq{roro1} is unimportant for the EL equation where it can be subsumed into the normalization factor $\lambda$, but must be retained when computing the worm-like chain free energy. Then, consistency requires that
\beq
\xi \approx  \frac{1}{3 \ell_{p}}  \left  ( \frac{3}{2}  \tilde{a}_{r} ^{2}  - 3 \tilde{a}_{r}  \right )
\eeq 
where the chain persistence length $\ell_{p}$ should be interpreted in units of the rod length $L_{r}$.  From the above the dependence of $\xi$ on the bare alignment amplitude $a_{r}$ is easily resolved and we find:
 \beq
 \xi \approx 1+ \ell_{p} +   |a_{r} |  -\sqrt{(1 +  \ell_{p})^{2} + 2 | a_{r} |  \ell_{p} }, \hspace{0.3cm} (a_{r} \ll 0 )
 \label{xi1}
 \eeq 
The correction factor vanishes in the rigid rod limit, $\lim_{\ell_{p} \rightarrow \infty} \xi = 0$, as it should.

\subsection*{Nematic polymers}

We may repeat the analysis for the case of conventional nematic polymers as encountered in the polymer-dominated $N^{+}$  phase using a slightly different route. For $a_{r} \gg 1$ the average polar deflection angle will be small and we may expand the worm-like chain term up to quadratic order in $\theta$. Using the asymptotic relation $\pp(t) \sim 1- 3\theta^{2}/2 $ and ignoring any constant factors we find a simple approximation valid for $|t|$ close to unity (strong alignment): 
\begin{align}
\frac{\nabla^{2}  \rho_{r}^{1/2}}{  \rho_{r} ^{1/2} }  \sim - \frac{3}{2} (  \tilde{a}_{r} ^{2}  + 2  \tilde{a}_{r} ) \pp(t) 
\label{roro2}
\end{align}
Then, in analogy with the preceding case we find an expression identical to \eq{xi1} except for a minus sign:
\beq
 -\xi \approx 1 + \ell_{p} + |a_{r}|  - \sqrt{(1 +  \ell_{p})^{2} + 2 |a_{r}|  \ell_{p}}, \hspace{0.3cm} (a_{r}  \gg 0)
\label{xi2}
\eeq
 This simple scaling result confirms our expectation, namely that a small degree of backbone flexibility leads to a reduction of the alignment amplitude for the polymers, since $| a_{r} + \xi | $ is always smaller than $| a_{r}  |$. For strongly aligned systems, this effect turns out to be of equal strength for both nematic and anti-nematically ordered polymers.

\section*{Appendix B: Stability of biaxial nematic order}

\begin{figure*}[ht]
  \includegraphics[width=.8 \textwidth]{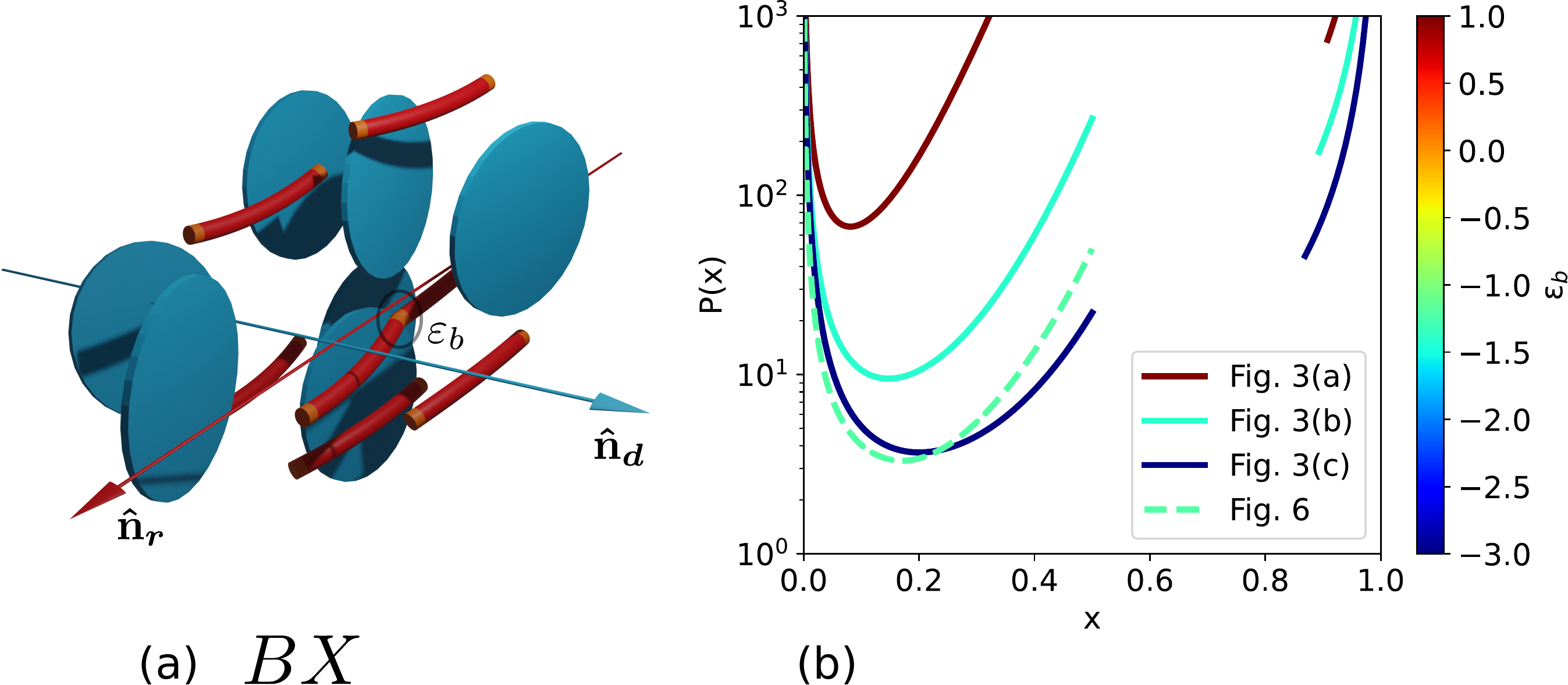}
\caption{(a) Schematic representation of the biaxial nematic phase $BX$ with orthorhombic ($D_{2h}$) point-group  symmetry, characterized by mutually perpendicular nematic directors for each species ($\bm{\hat{\textnormal{\bfseries n}}_r}$ for the polymers and $\bm{\hat{\textnormal{\bfseries n}}_d}$ for discs). (b) Bifurcation curves locating the transition from uniaxial (U) order at low osmotic pressure $P$ to biaxial (BX) nematic order at high pressure for a number of relevant cases studied in the main text. Here, $x$ denotes the disc mole fraction.  }
\label{fig:bx}
\end{figure*}

So far, we have overlook the possibility of biaxial nematic order in which both polymers and discs order along mutually perpendicular directors (see \fig{fig:bx}(a)).
In order to tentatively locate the transition toward biaxial ($BX$) nematic order, we  apply a simple bifurcation analysis in which we probe the stability a uniaxial ($U$) fluid against weakly biaxial fluctuations \cite{kayser,stroobants1984}. If the $U$-$BX$ transition is second-order, the bifurcation point at which biaxial solution emerge from the EL equations should pinpoint the actual phase transition.  We begin by generalizing the second-polynomial expansion \eq{p2expansion} to include biaxial nematic order using the addition theorem for spherical harmonics:
\begin{align}
\pp( \cos \gamma ) &= \pp(\cos \theta) \pp(\cos \theta^{\prime})  \nonumber \\ 
& + \frac{1}{12} \pp^{2}(\cos \theta) \pp^{2}(\cos \theta^{\prime}) \cos 2 (\varphi - \varphi^{\prime})
\end{align}
where $\varphi$ is the azimuthal angle describing the particle orientation  with respect a secondary director $\bn_{\perp} \perp \bn$. The coupled EL equations then attain an additional term that accounts for biaxiality: 
\begin{align}
&  \ell^{-1} \ln [  \rho_{r} (\ell, \oma ) \ell^{-1} ]   = \lambda_{r}    + \varepsilon_{b}  +  \alpha_{r}  \pp(\oma ) \nonumber \\ 
& +  \beta_{r } D(\oma)   +  \frac{L_{r}}{3\ell_{p}} \frac{\nabla^{2} [ \rho_{r} (\ell , \oma)]^{1/2}}{ [ \rho_{r} (\ell , \oma)]^{1/2} }
\label{lnrhorbiax}
\end{align}
and
\begin{align}
\ln [  \rho_{d} (\oma )] & = \lambda_{d}  + \alpha_{d} \pp(\oma)  + \beta_{d} D(\oma )  
\label{lndbiax}
\end{align}
in terms of the  weight function $D(\oma) = \sin^{2} \theta  \cos 2 \varphi $ and amplitudes: 
\begin{align}
\beta_{r} &=  \frac{15 }{16}  ( \rho_{r0} \bar{\Delta}_{r}  - 2q  \rho_{d0}  \Delta_{d} )  \nonumber \\
\beta_{d} &=   \frac{15 }{16}  (z \rho_{d0}  \Delta_{d}  - 2 q \rho_{r0}  \bar{\Delta}_{r} )
\label{alphabetabiax}
\end{align}
which feature the biaxial nematic order parameter of each species:
\begin{align}
\Delta_{r \ell} &= \rho_{r \ell}^{-1} \int d \oma \rho_{r}(\ell, \oma) D (\oma) \nonumber \\
\Delta_{d} &= \rho_{d0}^{-1} \int d \oma \rho_{d}( \oma) D (\oma ) 
\label{dr}
\end{align}
Similar to the uniaxial case the bar denotes a molecular-weight average according to $ \bar{\Delta}_{r} = \rho_{r0}^{-1}  \sum_{\ell} \rho_{r\ell} \Delta_{r \ell}$. 
Substituting the EL equations  into the  biaxial nematic order parameters and linearize for weakly biaxial amplitudes $|\beta| \ll 1$ we establish the condition under which a biaxial solution for the orientation distribution bifurcates from the uniaxial  one:
\begin{align}
\Delta_{r \ell} &= \rho_{r \ell}^{-1} \ell \beta_{r} \int d \oma \rho_{r }^{(U)} ( \ell, \oma) D^{2} (\oma) \nonumber \\
\Delta_{d} &= \rho_{d0}^{-1}  \beta_{d} \int d \oma \rho_{d }^{(U)} (\oma) D^{2} (\oma)
\label{brod}
\end{align} 
This linear criterion basically stipulates the conditions (overall particle concentration, composition and effective temperature) at which the  uniaxial nematic state is no longer guaranteed to be a local minimum in the free energy. Within the factorization Ansatz \eq{pfac} for the uniaxial molecular-weight distributions the condition simplifies into:
\begin{align}
\bar{\Delta}_{r} &=  \beta_{r} (2 m-1)  \langle D^{2}  (\oma) \rangle_{f_{G}} \nonumber \\
\Delta_{d} &=   \beta_{d}  \langle D^{2}  (\oma) \rangle_{f_{G}}
\label{brodgauss}
\end{align} 
The brackets denote an average according to the nematic or anti-nematic Gaussians  $f_{G}$ specified in \eq{gaussians}. Similarly, $m$ denotes the mean aggregation number of either the $N^{+}$ or the $N^{-}$ phase.
The averages are easily obtained in the asymptotic angular limits ($\theta \ll 1$ or $\psi \ll 1$) and the leading order contributions read:
\beq
\langle D^{2}  (\oma) \rangle_{f_{G}}  \sim 
 \begin{cases}
 4 /[ \alpha^{(1)}]^{2} \hspace{0.5cm}  \textrm{nematic}  \\
1/2 \hspace{1.2cm}   \textrm{anti-nematic} 
\end{cases}
\eeq 
 The $U$-$BX$ bifurcation condition \eq{brodgauss} is equivalent to the matrix equation ${\bf M} \cdot {\bf \Delta}  = \lambda_{e} {\bf \Delta} $ with ${\bf \Delta} = ( \bar{\Delta}_{r} , \Delta_{d}) $ and ${\bf M}$ given by the prefactors.  The eigenvalues $\lambda_{e} $ of the matrix ${\bf M}$ are required to be unity $(\lambda_{e} =1)$. The solution is:
 \begin{align}
 1 &= -\frac{15}{32} c w_{1} \left[ (1-x) + xzW - \right . \nonumber \\ 
 & \left . \sqrt{(1-x)^{2} + 2Wx(1-x)(8 q^{2} -z) +W^{2} x^{2}z^{2} }  \right]
\end{align}
with
\beq
w_{1} = 
 \begin{cases}
 (2m_{N^{+}}-1) 4/\alpha_{r}^{2}  \hspace{0.5cm}  N^{+}-BX  \\
 (2m_{N^{-}}-1)/2 \hspace{0.7cm}  N^{-}-BX
\end{cases}
\eeq 
and
\beq
W = 
 \begin{cases}
 (2m_{N^{+}}-1)^{-1} \alpha_{r}^{2}/8  \hspace{0.5cm}  N^{+}-BX  \\
 (2m_{N^{-}}-1)^{-1} 8/\alpha_{d}^{2} \hspace{0.5cm}  N^{-}-BX
\end{cases}
\eeq 

The numerical solutions are shown in \fig{fig:bx}(b).  The $N^{-}$-$BX$ solutions ceases to be internally consistent with the Gaussian approximation at  $x< 0.8$ given that the nematic order parameter $\alpha_{d}$ of the discs tends to get too low. No such inconsistency occurs for the $N^{+}$-$BX$ branches. In general, we find that the transition occurs at pressures that are beyond the ranges explored in the phase diagrams of the main text. The only exceptions are \fig{fig:L10}(c) and \fig{fig:qp} where the $N^{+}$ phase remains stable up to fairly large disc mole fractions and the $N^{+}$-$BX$ bifurcations are located within the monophasic $N^{+}$ regions (in red). The tentative conclusion from this analysis is in line with previous reports in literature \cite{mederos_overview2014}, namely that the stability of $BX$ nematic order is intimately linked to the excluded-volume asymmetry of the constituents which, in our case, is temperature-controlled.  Lowering the temperature reduces the typical asymmetry which then stabilizes well-mixed rod-disc nematics that subsequently may develop $BX$ order. We further note that disc-rich $BX$ phases seem much harder to stabilize than polymer-dominated ones as the $N^{-}$-$BX$ branches  generally do not intersect the small monophasic $N^{-}$ domains in the phase diagrams shown in the main text.

\section*{Data Availability}

The data that support the findings of this study are available from the corresponding author
upon reasonable request.

\bibliography{pub}

\end{document}